\documentclass[letterpaper,twocolumn,10pt]{article}
\usepackage{usenix}
\usepackage{booktabs}
\usepackage{xcolor}
\usepackage{pdfpages}
\usepackage{xspace}
\usepackage[dvipsnames]{xcolor}

\newcounter{notesec}[section]
\newcommand{\thenoteee}{\thesection.\arabic{notesec}}
\newcommand{\yue}[1]{\refstepcounter{notesec}{\bf\textcolor{blue}{$\ll$Yue~\thenoteee: {\sf #1}$\gg$}}}

\newcommand{\asher}[1]{\refstepcounter{notesec}{\bf\textcolor{brown}{$\ll$Asher~\thenoteee: {\sf #1}$\gg$}}}

\newif\ifsubmit
\submitfalse

\ifsubmit
    \newcommand{\needsref}[0]{}
\else
    \newcommand{\needsref}[0]{\textcolor{red}{[Ref?] }}
\fi

\usepackage{tikz}
\usepackage{amsmath}
\usepackage{pifont}
\usepackage{enumitem}
\usepackage{multirow}
\usepackage{longtable}
\usepackage{booktabs}
\usepackage{array}
\usepackage[margin=1in]{geometry}
\usepackage[most]{tcolorbox}
\usepackage{filecontents}
\usepackage{adjustbox}
\usepackage[table]{xcolor}
\usepackage[numbers,sort&compress]{natbib}

\newcounter{fcounter}
\newcommand{\finding}[1]{\refstepcounter{fcounter}
\vspace{0.25em}\noindent
\begin{tcolorbox}[
  colback=gray!5,        
  colframe=black!40,     
  boxrule=0.5pt,         
  arc=2pt,               
  left=2pt,
  right=2pt,
  top=2pt,
  bottom=2pt
]{%
\parbox{1\linewidth}{%
  \vspace{0.3em}{\bf
  {Finding~\arabic{fcounter}~(\fnumber{\arabic{fcounter}})}~--} {#1}
\vspace{0.3em}
  }
}
\end{tcolorbox}
\vspace{0.25em}
}

\newcommand\fnumber[1]{{$\mathcal{F}_{#1}$}}

\newtcolorbox{takeaway}[1]{
    lower separated=false,
    colback=blue!2!white,
    colframe=teal!90!black,
    fonttitle=\sffamily\bfseries,
    left=1pt,
    right=1pt,
    bottom=1pt,
    top=4pt,
    colbacktitle=teal!90!black,
    coltitle=blue!2!white,
    enhanced,
    attach boxed title to top left={yshift=-0.1in,xshift=0.15in},
    boxed title style={boxrule=0pt,colframe=white,},
    title={\color{white}{#1}}
}
\newcommand{\ignore}[1]{}
\newcommand\agentName{\texttt{CTFriend{}}}

\begin{document}

\date{}


\title{\Large \bf Understanding Human-AI Collaboration in Cybersecurity Competitions}


\author{
\parbox{\textwidth}{\centering
\textbf{Tingxuan Tang\textsuperscript{1}},
\textbf{Nicolas Janis\textsuperscript{1}},
\textbf{Kalyn Asher Montague\textsuperscript{1}},
\textbf{Kevin Eykholt\textsuperscript{2}},
\textbf{Dhilung Kirat\textsuperscript{2}},
\textbf{Youngja Park\textsuperscript{2}},
\textbf{Jiyong Jang\textsuperscript{2}},
\textbf{Adwait Nadkarni\textsuperscript{1}},
\textbf{Yue Xiao\textsuperscript{1}}\\[1.2ex]
\textsuperscript{1}William \& Mary,
\textsuperscript{2}IBM Research
}
}


\maketitle

\begin{abstract}
Capture-the-Flag (CTF) competitions are increasingly becoming a testbed for evaluating AI capabilities at solving security tasks, due to their controlled environments and objective success criteria.
Existing evaluations have focused on how successful AI is at solving individual CTF challenges in isolation from human CTF players. As AI usage increases in both academic and industrial settings, it is equally likely that human CTF players may collaborate with AI agents to solve CTF challenges.
This possibility exposes a key knowledge gap: {\em how do human players perceive AI CTF assistance; when assistance is provided, in what ways do they collaborate and is it effective with respect to human performance; how do humans assisted by AI compare to the performance of fully autonomous AI agents on the same set of challenges}.
%
We address this gap with the first empirical study of AI assistance in a live, onsite CTF.
In a study with 41 participants (out of the total 95 that participated in the CTF), we qualitatively study {\sf (i)} how participants’ perception, trust, and expectations shift before versus after hands-on AI use, and {\sf (ii)} how participants collaborate with an instrumented AI assistant. 
Moreover, we also {\sf (iii)} benchmark four autonomous CTF agents on the same fresh challenge set to compare outcomes with human teams and analyze agent trajectories.
We find that, as the competition progresses, teams 
increasingly delegate larger subtasks to the AI, giving it more agency.
Interestingly, CTF challenges solving rates are often constrained not by the model's reasoning capabilities, but rather by the human players: ineffective prompting and poor context specification become the primary bottleneck. 
Remarkably, autonomous agents that self-direct their prompting and tool use bypass this bottleneck and outperform most human teams, coming in second overall in the CTF competition.
We conclude with implications for the future design of CTF competitions and for building effective human-in-the-loop AI systems for security.

%
%
%

\end{abstract}

\section{Introduction}

Large language models (LLMs) are improving at an exceptional pace and are increasingly used for security tasks, from vulnerability discovery~\cite{chatgpt_vulnerable_manage_2024, huynh2025detecting, nong2024chain, yu2024insight,sheng2024lprotector, lin2025llm, li2025vulsolver,weissberg2025llm} and exploit generation~\cite{zhang2023well,wu2023deceptprompt, fang2024llm, huang2023catastrophic,chen2025exp} to patch development~\cite{kim2025logs, yu2024llm, kulsum2024case, xu2025revisiting, zhang2025patch}.
Capture The Flag (CTF) competitions have emerged as the de facto benchmark for evaluating the capabilities of AI agents in security-related challenges. CTF challenges are designed to simulate realistic vulnerability scenarios that require reconnaissance, hypothesis formation, iterative testing, and exploitation. CTFs also define clear success criteria (i.e., submitting the flag) and can be hosted in controlled environments for reproducible evaluation.

Recent work on AI for CTF has largely focused on \textit{AI capability-centric} evaluations, including benchmark construction~\cite{shao2024nyuctfbench,yang2023language, buckley2021regulating}, AI agent framework design~\cite{zou2026ctfagent,abramovichenigma,mayoral2025cybersecurity,abramovich2025enigma,shao2025craken,zhang2025cybench}, and training pipelines that improve end-to-end performance~\cite{zhuo2025training}.
While these efforts quantify what AI systems can do in isolation, they offer limited insight into the potential of \textit{human-AI collaboration}.
Such collaboration is necessary as full automation of security workflows remains challenging. Security-relevant tasks usually require contextual judgment and iterative verification.
Over-delegating security tasks to AI introduces practical risks, including hallucinated conclusions~\cite{spracklenpackage}, unintended tool misuse~\cite{mousavi2024investigation}, wasted analyst time, and computational resources.
In this paper, we study how participants perceive and leverage AI in a security context and the effectiveness of this collaboration. 
We address these questions through an in-person  CTF competition, which provides a measurable and ethically controlled environment for observing human-AI interaction under realistic time pressure while solving security challenges.
Specifically, we study the following three research questions:


%
%
%


\noindent\textbf{\textit{RQ$_1$: What expectations do participants have for AI, specifically in a CTF scenario?}} How do perception, trust, and expectation shift? How does participants' AI expertise impact their CTF scores?

\noindent$\bullet$\textit{ User survey analysis of common perception (RQ$_1$).}
We performed a pre-survey and post-survey with \textbf{41} CTF participants to measure participant expectations and reflections.
We find that participants' expected AI effectiveness and willingness to use AI in future competitions decreased after hands-on use ($\mathcal{F}_{1}$). Participants attributed this shift to recurring model failures, including flawed reasoning, hallucinations, and non-working code ($\mathcal{F}_{2}$), which in turn reduced trust in AI outputs, especially among participants with higher CTF domain knowledge. Finally, we observe that AI expertise can partially compensate for limited CTF domain knowledge: participants with low CTF expertise but high AI expertise achieved competitive scores, whereas participants with intermediate CTF expertise but novice AI expertise solved few challenges ($\mathcal{F}_{3}$), underscoring effective AI use as a key determinant of performance.

\noindent\textbf{\textit{RQ$_2$: How do humans actually collaborate with an AI assistant during a live CTF?}} What distinguishes effective collaboration from ineffective collaboration? Do collaboration strategies shift in a competition environment?

\noindent$\bullet$\textit{ Qualitative analysis of Human-AI interactions (RQ$_2$)}
To understand how CTF players actually collaborate with an AI assistant during a live competition, we develop and deploy an instrumented assistant, \agentName{}, for participant to interact with and qualitatively analyze \textbf{2,299} chat messages to identify emergent interaction patterns, effective strategies, common failure modes, and how human-AI \textit{leadership} (i.e., who drives the next step) evolves over time. 
We find that participants' stated intention to work \textit{cooperatively} with the assistant often collapses under real-time competitive pressure: many teams shift toward end-to-end delegation, letting the AI take on entire tasks rather than using it for incremental support ($\mathcal{F}_{6}$).
Whether this delegation helps depends strongly on expertise. High-expertise users tend to employ higher-quality prompts with richer technical details and more effective prompt engineering strategies to achieve better outcomes while
novices used low-quality prompts, leading to errors and
task failures ($\mathcal{F}_{7}$).
Lower-expertise users were susceptible to low-risk, low-success rate strategies and adopted \texttt{"answer shopping"} through of repetitive prompting ($\mathcal{F}_{8}$).
This divergence is not inevitable. We observed that some participants with zero or limited CTF knowledge were able to use the AI as a constructive learning scaffold that allowed them to rapidly acquire missing domain concepts and solve challenges they would otherwise be unable to approach ($\mathcal{F}_{9}$).

\noindent\textbf{\textit{RQ$_3$:How do autonomous CTF agents compare to humans on the same set of challenges?}} Can autonomous agents outperform humans, and what are the limits?

\noindent$\bullet$\textit{ Empirical comparison between Human and Autonomous AI agents  (RQ$_3$)}
We evaluate four autonomous agent frameworks paired with three Claude-family models (12 configurations total) on the same fresh challenge set used in the live competition, and compare their performance against human teams.
We find that advanced agents can outperform most human teams while requiring only $\sim$1/5 of the cumulative runtime, with the strongest configuration reaching \texttt{4900} points (second among the top-10 human teams) at a cost of \$96.32 in API usage ($\mathcal{F}_{11}$).
Performance varies sharply across agent designs: agents that combine long-horizon planning with robust interactive tool support consistently perform best, whereas restricted tool wrappers and fixed tool sets correlate with early plateaus ($\mathcal{F}_{12}$).
At the same time, backbone model capability sets the ceiling; under weaker models, different frameworks converge to similarly low performance ($\mathcal{F}_{13}$).
We also find that challenges that are difficult for humans are not necessarily difficult for agents, and vice versa.
Some challenges that are operationally hard for agents (e.g., requiring intensive environment interaction) are comparatively manageable for humans, and vice versa ($\mathcal{F}_{14}$).
This mismatch motivates human-in-the-loop \textit{pair hacking}, where the agent runs autonomously for high-throughput work while humans provide sparse steering and verification to overcome brittleness ($\mathcal{F}_{15}$).

\section{Background}
\label{sec:background_ctf}
CTF competitions are a widely used educational tool to help students, practitioners, and security enthusiasts demonstrate and polish their skills at solving security challenges.
CTF challenges cover a broad spectrum of security domains such as cryptography, reverse engineering, web exploitation, and forensics. They are inspired by real security vulnerabilities, but each CTF competition varies in challenge design. It is rare that the exact same solution for one CTF challenge can be re-used in a future CTF. These factors make CTF competitions a valuable testbed for investigating human-AI collaboration. 


To elaborate, most CTF competitions are time-limited and follow a Jeopardy-style format. The same set of challenges is provided teams that contains a challenge description and challenge artifacts (e.g., binaries, source code, packet traces, disk images, etc.). Participants are expected to analyze challenge artifacts, potentially interact with live sandboxed services (e.g., a vulnerable web or network endpoint), and identify the relevant weakness which will lead to solving the challenge. A partcipant prove they have solved the challenge by recovering the ``flag'', often a secret token or text string hidden in the challenge, and submitting it.
Upon solving a challenge, participants gain the points associated with the estimated difficulty tier of the challenge (e.g., easy, medium, and hard challenges). 
At the end of the CTF, the team with the most points wins, with time-to-solve being used as a common tie-breaker if needed.


CTFs can be further tailored to scale in difficult according to the target audience. K-12-oriented CTFs are considered to be the easiest challenges leveraging well-known security vulnerabilities with only a few exploitation steps (e.g., picoCTF~\cite{picoctf2026}), followed by university-level CTFs that represent moderate difficulty (e.g., the CSAW CTF~\cite{csawctf2025}), followed by expert-level CTFs (e.g., the DEF CON CTF~\cite{defconctf}).
As we describe in detail in Section~\ref{sec:liveCTF}, this study focuses on an in-person university-level CTF that falls on the higher end of the difficulty spectrum, i.e., it has a larger number of expert-level challenges than a typical university CTF.

%

\section{Related Work}
\label{sec:relwork}
Software systems are growing in size and connectivity, which increases the attack surface and the cost of manual security analysis~\cite{zhuo2025training}. 
To scale security analysis beyond what humans can handle alone, programs such as the DARPA Cyber Grand Challenge and the DARPA AI Cyber Challenge (AIxCC) have accelerated interest in using AI, particularly LLMs/AI-agents, to help detect, exploit, and fix software flaws~\cite{song2015darpa,DARPA}.
In this context, CTF competitions have emerged as the de facto benchmark for evaluating the capabilities of AI agents in cybersecurity tasks, as CTFs offer diverse security tasks in a controlled environment and have clear success criteria. 
This shift is also reflected in the emergence of AI-first CTF competitions, such as Hack The Box's \textit{Neurogrid CTF}, which explicitly organized around deploying AI agents with MCP integration for challenge solving~\cite{neurogridctf}.

Recent work on leveraging AI for CTFs has progressed along three directions: benchmark construction, autonomous agent design, and training pipelines. 
On the benchmarking side, NYU CTFBench curates 200 challenges from prior CSAW CTF competitions at roughly university-level difficulty~\cite{shao2024nyuctfbench}. 
Intercode-CTF provides 100 problems collected from PicoCTF, a large-scale security competition aimed at high-school-level participants~\cite{yang2023language}. 
%
CTFKnow~\cite{ji2025measuring} measures CTF-relevant technical knowledge using thousands of multiple-choice and open-ended questions.

Beyond static benchmarks, researchers have built autonomous CTF agents that couple LLM reasoning with tools, memory, and iterative environment interaction (e.g., ENIGMA, CRAKEN, and Cybench)~\cite{abramovich2025enigma,shao2025craken,zhang2025cybench}. 
These agents run in a sandbox with access to challenge artifacts and iteratively follow a plan–act–observe loop (reasoning about the steps, executing commands (e.g., Bash/Python, tool calls), and observing outputs as feedback for the next step) until they submit a valid flag or exhaust their allotted turns or resources.
To further improve agent performance, training platforms such as CTF-DOJO collect and synthesize executable agent trajectories in realistic cybersecurity environments to support LLM training or fine-tuning on offensive cybersecurity tasks~\cite{zhuo2025training}.
Our work is different from prior CTF studies by focusing on how humans use AI in practice during live problem-solving.
This lens matters because real-world impact depends not only on AI capability in isolation, but also on how practitioners integrate AI into workflows, including what they delegate, how they provide context, how they verify outputs, and how they recover from errors. By combining a live CTF study of human perceptions and interaction logs with a controlled comparison of humans and autonomous agents on the same challenges, we provide empirical insights for designing and evaluating effective human-in-the-loop AI systems for security.

\section{Experimental Setting} 
\begin{figure}[t]
  \centering
  \includegraphics[width=1.0\linewidth]{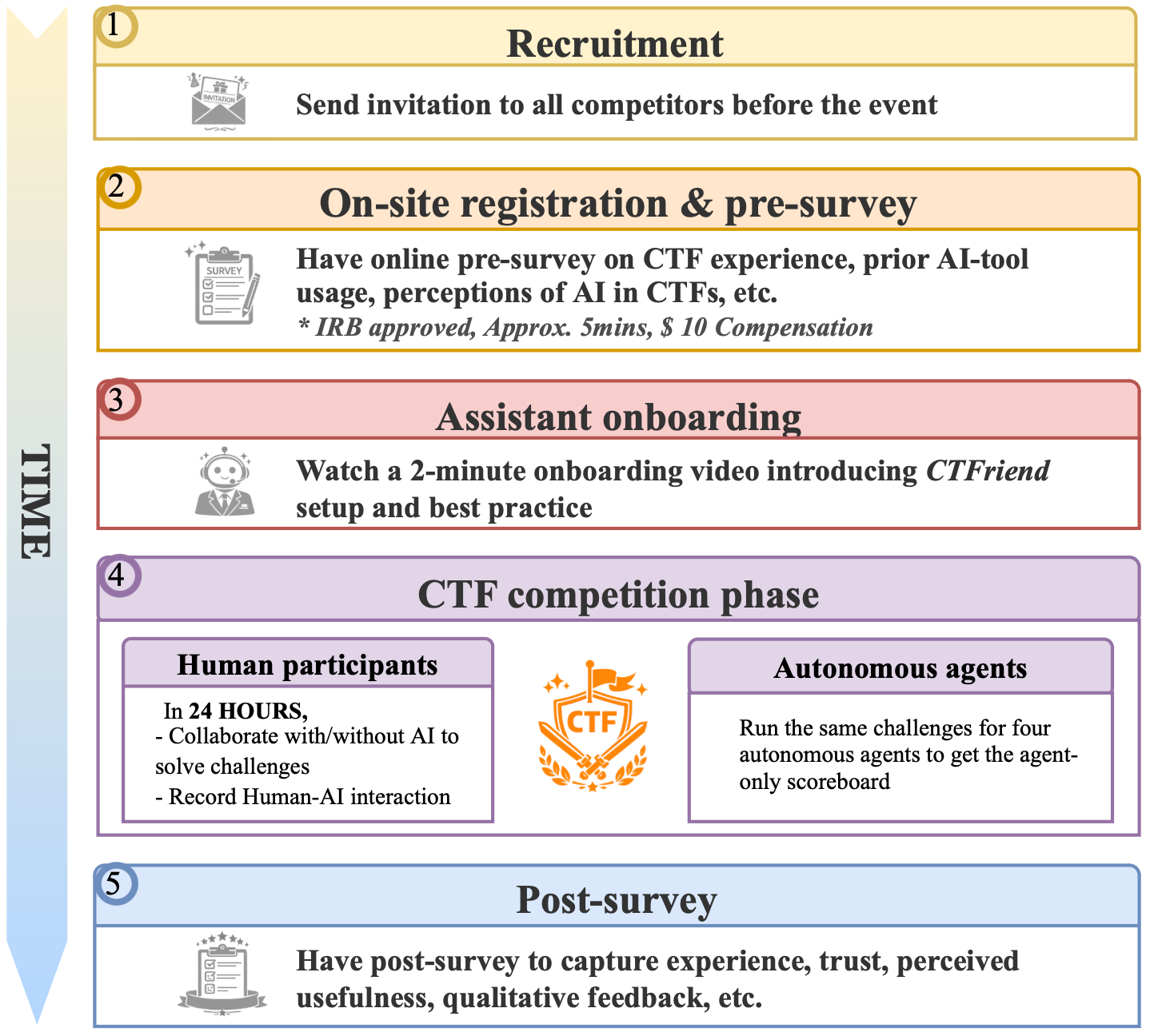}
  \caption{User Participation Workflow} 
  \label{fig: study_workflow}
\end{figure}

\subsection{A Live University-level CTF Event}
\label{sec:liveCTF}
We hosted a live, on-site CTF competition on campus and conducted our experiments in this setting. 
The event included 17 newly designed challenges spanning 5 CTF categories, including forensics, cryptography, reverse engineering, web exploitation, and others. Challenges were organized into three difficulty tiers: \texttt{easy}, \texttt{medium}, and \texttt{hard}, worth \texttt{300}, \texttt{500}, and \texttt{1000} points respectively, for a total of \texttt{7700} available points. Table~\ref{tab: challenge_stats} summarizes the distribution. 
This was a university-level competition: most challenges were comparable to (and slightly harder than) typical CSAW-level tasks, with 1-2 challenges closer to picoCTF-level and 2-3 challenges approaching DEFCON CTF-level difficulty. All challenges were authored specifically for this event with support from a security firm and were not previously released, reducing the risk of contamination from existing writeups.
%

\subsection{Study Procedure}
This section describes the end-to-end study workflow (Figure~\ref{fig: study_workflow}). For each participant, the study lasted approximately 24.5 hours and consisted of: a 5-minute informed consent process, a 2-minute AI-assistant usage training video, the live CTF competition phase (24 h), and a 20-minute post-study survey. 
All procedures were approved by our institution's Institutional Review Board (IRB). Participants received \$10 compensation for completing the study. 
Our study consists of the following five steps:

\vspace{2pt}\noindent\textbf{\ding{202} Recruitment.} Prior to the event, we sent an invitation to all competition participants to ask whether they were willing to take part in the study. We recruited 41 participants, with demographics described in \S~\ref{sec:demographics}. 

\vspace{2pt}\noindent\textbf{\ding{203} Onsite registration and pre-survey.} Before the game began, we set up a registration desk where interested participants enrolled in the study. Participants first completed an online pre-survey (approximately 5 minutes) covering CTF experience, prior AI-tool usage, and initial perceptions of AI assistance in CTFs.

\vspace{2pt}\noindent\textbf{\ding{204} Assistant onboarding.} After the pre-survey, we created an account for each participant to access our CTF AI assistant \agentName{} (Appendix~\S\ref{sec:implementation}). The assistant provided access to a range of open-source and commercial language models (details in Appendix~\S\ref{sec:implementation}). We also provided a \textbf{2}-minute onboarding video demonstrating the interface and expected usage of \agentName{}.

\vspace{2pt}\noindent\textbf{\ding{205} CTF Competition phase.} The CTF competition ran for \textit{24 hours}. Participants collaborated with AI to solve challenges. Our backend recorded all human-AI interactions (e.g., prompts and model responses), enabling our log-based analysis of collaboration in \textbf{RQ$_2$} (\S~\ref{sec:rq2}).

\textit{Parallel autonomous-agent evaluation.} In parallel with the human competition, we ran multiple autonomous AI agents on the same challenge set to produce an agent-only scoreboard. These experiments support \textbf{RQ$_3$} by enabling a direct comparison between human and autonomous performance on the same fresh challenges (\S~\ref{rq3}).

\vspace{2pt}\noindent\textbf{\ding{206} Post-survey.} After the competition ended, we distributed a post-CTF survey to all participants to capture their experiences using AI assistance during the event, including perceived value, trust, and feedback. These responses support our survey-based analysis in \textbf{RQ$_1$} (\S~\ref{sec:rq1}).




\section{RQ$_\textbf{1}$: Users' Perception of AI in CTF} 
\label{sec:rq1}
 RQ$_\textbf{1}$ examines how participants perceive AI assistance in CTFs (\S~\ref{sec:reflections}) and analyzes how their performance varies in the context of their backgrounds (\S~\ref{sec:CTF_performance}). 

\subsection{Methodology}
\label{sec:methodology}
Our survey design is guided by key questions that intuitively emerge when we consider user perceptions of AI assistance in security-oriented tasks (see details in \S~\ref{Survey Design}). For instance, what do participants expect AI tools to accomplish in a CTF setting, and how do these expectations differ after the competition? What aspects of interacting with an AI assistant do participants find helpful versus frustrating during a competition? After hands-on exposure, do participants intend to incorporate AI assistance in future CTFs or broader security workflows?

%




\paragraph{Participants' Demographics} 
\label{sec:demographics}
\begin{table}[t]
\centering
\small
\setlength{\tabcolsep}{6pt}
\renewcommand{\arraystretch}{1.15}

\begin{adjustbox}{width=\linewidth}
\begin{tabular}{lrrr lrrr}

\toprule
\multicolumn{4}{l}{\textbf{(1) Grade}} & \multicolumn{4}{l}{\textbf{(2) Major}} \\
\cmidrule(lr){1-4}\cmidrule(lr){5-8}
 & \textbf{N} & \textbf{\%} &  &  & \textbf{N} & \textbf{\%} &  \\
\midrule
\rowcolor{black!8}
Undergraduate         & 34 & 82.9 &  & Computer Science                   & 23 & 56.1 &  \\
Graduate              &  6 & 14.6 &  & Cybersecurity                      &  8 & 19.5 &  \\
\rowcolor{black!8}
K-12 / High School    &  1 &  2.4 &  & Info Sys. \& Engineering &  4 &  9.8 &  \\
                      &    &      &  & Other                               &  6 & 14.7 &  \\

\addlinespace[0.8em]
\multicolumn{4}{l}{\textbf{(3) CTF Domain Expertise}} & \multicolumn{4}{l}{\textbf{(4) AI Usage Expertise}} \\
\cmidrule(lr){1-4}\cmidrule(lr){5-8}
 & $\mathbf{E_S}$ & \textbf{N} & \textbf{\%} &  & $\mathbf{E_{AI}}$ & \textbf{N} & \textbf{\%} \\
\midrule
\rowcolor{black!8}
Zero Expertise    & 0-1 & 12 & 29.3 & Zero Expertise & 0 &  5 & 12.2     \\
Novice            & 1-2 & 10 & 24.4 & Novice        & (0,5] & 17 & 41.5  \\
\rowcolor{black!8}
Intermediate      & 2-3 & 16 & 39.0 & Intermediate  & (5,10] & 10 & 24.4  \\
Expert            & >3 &  3 &  7.3 & Expert        & >10 &  9 & 22.0    \\

\addlinespace[0.8em]
\multicolumn{8}{l}{\textbf{(5) Prior CTF Experience}} \\
\cmidrule(lr){1-8}

\textbf{\# Prior CTFs} & \textbf{N} & \textbf{\%} & \textbf{} &
\textbf{}             & \textbf{N} & \textbf{\%} & \textbf{} \\
\midrule

\rowcolor{black!8}
0     & 17 & 41.46 &  & 5 --- 8 & 3 & 7.32 &  \\
1 --- 4 & 19 & 46.34 &  & 8 +     & 2 & 4.88 &  \\

\end{tabular}
\end{adjustbox}
\caption{Participant demographics and expertise.}
\label{tab: participant_demo}
\end{table}

To provide a clearer understanding and categorization of the study participants (N=41), we asked participants to report their academic backgrounds, prior CTF experience, expertise in CTF subject domains. We then aggregated these scores to get CTF expertise score for each individual ($E_S$).  We also asked users to self-report their AI expertise in terms of number of security challenges solved with AI ($E_{AI}$. For analysis, we grouped participants as shown in Table \ref{tab: participant_demo}.  Generally, we consider zero-expertise and novice participants to have \texttt{low expertise}, and intermediate and expert participants to have \texttt{high expertise}.

\subsection{Results: Participants' Reflections}
\label{sec:reflections}

\paragraph{Participant Expectations vs. Reflections}
To evaluate participants' expectations, beliefs, and understanding of AI, we have analyzed the data gathered in the pre-CTF survey by demographic group.  We then compare each analysis point to the participants' reflection data gathered in the post-CTF survey.  Thus, we can observe how different demographics perceived AI use in security scenarios before the competition, and how those understandings changed after experiencing the live event.

\noindent$\bullet$\textit{ Expectation, trust, and attitude toward AI use:}
We measure how participants’ perceptions of AI shift after hands-on use during the onsite CTF.
Overall, participants initially overestimated how much the AI would improve their ability to solve challenges; after the event, expectations for AI solve counts dropped by $\Delta 0.47\downarrow$, which aligns with the ineffective collaboration behaviors we later observe in logs (e.g., over-delegation and answer shopping; see RQ$_2$).
Participants' also reported that suggested steps were often difficult to follow under time pressure and occasionally contained hallucinated or misleading details (see $\mathcal{F}_2$).
Overall, attitudes were split, with 44\%  of participants reporting they still plan to use AI in security, viewing it as a fast learning tool, whereas the others reported reluctant due to perceived performance limits and the risks of relying on AI outputs.

\finding{Following this CTF competition, participants' perceptions of AI performance on challenges and trust in AI output decreased.}

\noindent$\bullet$\textit{ Error and Frustrations:} The most frequently observed errors reported in the post-CTF survey were flawed reasoning and hallucinations, which most users reported as a time waste in the competition. Additionally, users reported errors due to AI ethical constraints (i.e. safety constraints, content moderation, or value alignment), with several users singling it out as the most frustrating part of using AI assistance in the competition.  However, it is possible that many participants lacked full understanding of what constituted an error or why it occurred, as evidenced by autonomous agents (\S~\ref{sec:agents} not encountering the same issues.  This is discussed further in \S~\ref{par:strategies}.

\finding{Participants identified flawed reasoning, hallucinations, non-working code, and ethical guardrail restrictions as common AI errors that served as points of friction and time loss in the competition.}

\noindent$\bullet$\textit{ Understanding and Satisfaction:}
Similarly, participants reported that AI frequently failed to provide complete, correct and actionable solutions. As shown Figure \ref{fig: user_experience}, most participants rated the AI as moderate to low in its ability to generate complete, correct, and actionable solutions. However, participants also reported that AI demonstrated a strong ability to understand user inputs and provide clear and easy-to-understand responses.  It is important to note, however, that these actionability and completeness grades may be partially due to users' inability to implement the challenge solutions, which is discussed further in \S~\ref{par:strategies}.

\finding{Participants reported AI had solid understanding of user input and clarity in output, but solutions lacked in completeness and actionability.}

\subsection{Results: Expertise and Performance}
\label{sec:CTF_performance}
\paragraph{Competition Results}

To determine the significance of participants' backgrounds on competition performance, we analyzed users' AI and CTF expertise in conjunction with their individual scores from the competition. Discussion of individual score calculation is provided in Appendix~\S\ref{sec:qual_supplemental}.

\noindent$\bullet$\textit{ Prerequisites for Success:} Unsurprisingly, users who reported high CTF and AI expertise generally scored the highest in the competition. Interestingly, this pattern held regardless of prior CTF experience. 
For example, three of the top four and 8 of the top 10 highest-scoring individuals in the competition reported attending two or fewer prior CTFs, four of whom reported that this was their first CTF.  Meanwhile, all but one of these participants reported at least intermediate-level domain expertise, with the highest-scoring individuals having the highest domain knowledge of this subgroup.  For a specific example, User 9, who competed on a high-scoring, three-person team, reported the highest level of domain expertise and accounted for ~60\% of their team's final score despite having attended the fewest prior CTFs (1).

\finding{Participants who reported higher levels of AI and CTF domain expertise saw the highest individual scores in the competition among participants, even when having less direct CTF experience.}

While users who reported higher levels of AI and CTF expertise achieved the highest scores, 64\% of participants  who reported low CTF expertise but high AI expertise were still able to achieve intermediate to high scores in the competition, and 91\% completed multiple challenges.  This suggests that their AI expertise may have partially made up for their lack of domain expertise, highlighting the value of AI aid in CTFs.  In contrast, 85\% of participants who reported neither AI nor CTF expertise were only able to complete one or fewer challenges.  Interestingly, 2 of the 3 participants who reported at least intermediate CTF expertise but novice or lower AI expertise were only able to complete one or fewer challenges.

\finding{Participants who reported low CTF domain expertise but high AI expertise were able to achieve high scores in the competition, suggesting the value of AI in filling in user knowledge gaps.}

\section{RQ$_\textbf{2}$: Human-AI Collaboration}
\label{sec:rq2}
RQ$_1$ characterizes participants’ perceptions of AI assistance in CTFs. RQ$_2$ goes one step further by examining how participants \emph{actually} collaborate with AI during security-relevant problem solving. 
To address RQ$_2$, we developed and deployed an instrumented AI assistant, \agentName{}, to record human-AI interactions. In total, we collected \textbf{2,299} messages across \textbf{168} chat logs from \textbf{38} participants. We then qualitatively analyzed these logs, yielding \textbf{5} findings on \textit{emergent interaction patterns}, \textit{successful collaboration strategies}, and \textit{AI risk and reward}.





\subsection{Methodology}

\subsubsection{Design and Deployment of \agentName{}}

To study human–AI collaboration during a live CTF, we built and deployed \agentName{}, a web-based application made available to participants.
%
Importantly, \agentName{} is not intended to operate as an automatic CTF solver, but rather as an AI assistant, supporting participants in challenge solving by providing access to Claude-family models (Sonnet 4.5, Opus 4.1, and Haiku 3.5), while keeping flag submissions under human control and allowing researchers to monitor and collect human–AI interaction data. We release the \agentName{} code in~\cite{projectWebsite}.




\vspace{4pt}\noindent\textbf{System Overview.} 
Figure \ref{fig:agent_workflow} presents an overview of \agentName{}’s architecture. Specifically,
(\ding{202}) The Streamlit-based web UI serves as the primary interaction layer, providing a conversational interface that mirrors commonly used AI assistants and persistently displays conversation history to support iterative and multi-turn problem solving.
(\ding{203}) The AI agent layer orchestrates interactions with multiple LLMs through a unified interface, with API access managed by the backend, allowing participants to use diverse AI capabilities without providing their own credentials and reducing friction in time-constrained CTF settings.
(\ding{204}) To support CTF-specific problem solving, the agent is equipped with a modular MCP tool layer and a retrieval-augmented CTF knowledge base.  
(\ding{205}) All user interactions, conversation histories, and feedback signals are persistently stored in the database layer, supporting systematic analysis of human–AI interaction behaviors.
(\ding{206}) Finally, a dedicated monitoring and visualization stack provides real-time insights into system health and application usage, ensuring reliable operation during live competitions and comprehensive observability for empirical study. The implementation details are in Appendix~\ref{sec:implementation}.

\subsubsection{Qualitative Analysis}
\label{sec:quali_methodology}
We analyze \agentName{} interaction logs using a two-pronged qualitative approach: a codebook-based analysis to systematically identify recurring collaboration patterns and error modalities at scale, complemented by an expert case review of selected logs to assess domain-specific technical correctness of AI's responses and user comprehension.
%

\vspace{2pt}\noindent\textbf{Coding protocol.} To analyze a wide set of concepts, we adopted a hybrid inductive-deductive coding protocol.

\noindent $\bullet$ \textit{Code System Development: }   First, we developed deductive codes based on our survey questions and an existing taxonomy of AI errors and prompt engineering strategies.  We then performed inductive coding in two phases: first, we executed a thematic exploratory phase where our coder performed an initial coding of a randomly-selected subset of chat logs until saturation was reached (no new codes for 50 user-turns).  Next, we executed a confirmatory phase on 40 randomly-selected logs, where the coder refined content analysis subcodes and definitions based on the observed themes.

\noindent $\bullet$ \textit{Coding and Validation: } Upon completion of the exploratory phase, the codebook was frozen and main coding effort performed.  Finally, after 21-day washout interval, the coder performed a re-coding of a 50 randomly-selected logs, resulting in a Positive Agreement = 0.74 and median $J_i=.79$ with IQR = 0.17 on significant codes (i.e., codes used in analysis/findings).  All coding was performed blind to prior labels, hypotheses, and participant condition.  In total, this process took over 170 hours and uncovered 9312 occurrences of 72 codes and subcodes.

\noindent $\bullet$ \textit{Coding Unit: } We coded user prompt-specific behaviors at the user-turn level and AI response behaviors at the assistant-turn level. Other patterns were coded at the episode level, where an episode corresponds to a single user task. A definition/frequency codebook is provided in Appendix~\S\ref{sec:codebook}, with additional information in \ref{sec:quali_analysis}. The complete code system and protocols and can be found on the paper website~\cite{projectWebsite}.

\noindent $\bullet$ \textit{Code and Log Analysis: } Coding and analysis was performed using MaxQDA~\cite{maxqda}.  To account for variation in log length, we report prevalence as participant-normalized rates (median fraction of eligible turns/episodes per participant).  For calculations of co-occurrence or odds ratio, operationalized variables are provided with corresponding uncertainty metrics in Appendix~\S\ref{sec:quali_analysis}.  To deepen our analysis, we also invited the author of the CTF challenges to perform an expert review of six representative logs from intermediate and high-scoring teams and nine autonomous agent logs from ~\ref{sec:agents} to provide insight on the critical advantages that helped the best teams stand out in the competition ($\mathcal{F}_8$).

\subsection{Results: Interaction Behaviors}
\label{sec:competition_behavior}

We identified several notable interaction patterns between users and AI during our study, including some that demonstrate a likely change in behavior patterns due to the time-constrained nature of the CTF.

\vspace{2pt}\noindent\textbf{Emergent Interaction Patterns.}
In the pre-CTF survey, we asked participants to report their \textit{expected} collaboration patterns with AI. We then compared these self-reports against their real-world behavior observed through our qualitative coding of the interaction logs. 
This comparison reveals a clear gap between what participants believed they would do and what they actually did during the competition.
Specifically, while \textit{trial-and-error} was the most common interaction pattern reported by users (80\%) in the pre-CTF survey, \textit{delegation} was the dominant pattern that emerged during the competition.
Specifically, the most common observed strategy was a user providing the full challenge prompt along with a simple instruction such as \textit{``solve this''} or \textit{"how can I do this"}.
Additionally, although \textit{collaborative refinement}, \textit{confirmation-seeking}, and \textit{rejection of suggestions} (definitions in the codebook, Appendix~\S\ref{sec:codebook}) were reported at moderate rates in the pre-CTF survey (61\%, 63\%, and 41\%, respectively), the actual prevalence of these patterns was much lower (only 16\% on average, compared to 38\% for delegation). Both observations suggest a breakdown between participants' perspective on their interaction patterns and the empirical evidence.
%

\finding{Although participants expressed the intention to collaborate iteratively with AI,  in practice they more often delegated full tasks to the agent.}

\paragraph{Successful strategies and Failure mode.}
\label{par:strategies}

As discussed in \ref{sec:CTF_performance}, most participants who reported intermediate- or expert-level AI expertise achieved moderate-to-high scores in the competition, while the highest-scorers reported high expertise in both AI and CTF domains.  We observe that this driven primarily by two factors: prompting strategy and domain-knowledge injection.

\begin{figure*}[t]
  \centering
  \includegraphics[width=1.0\linewidth]{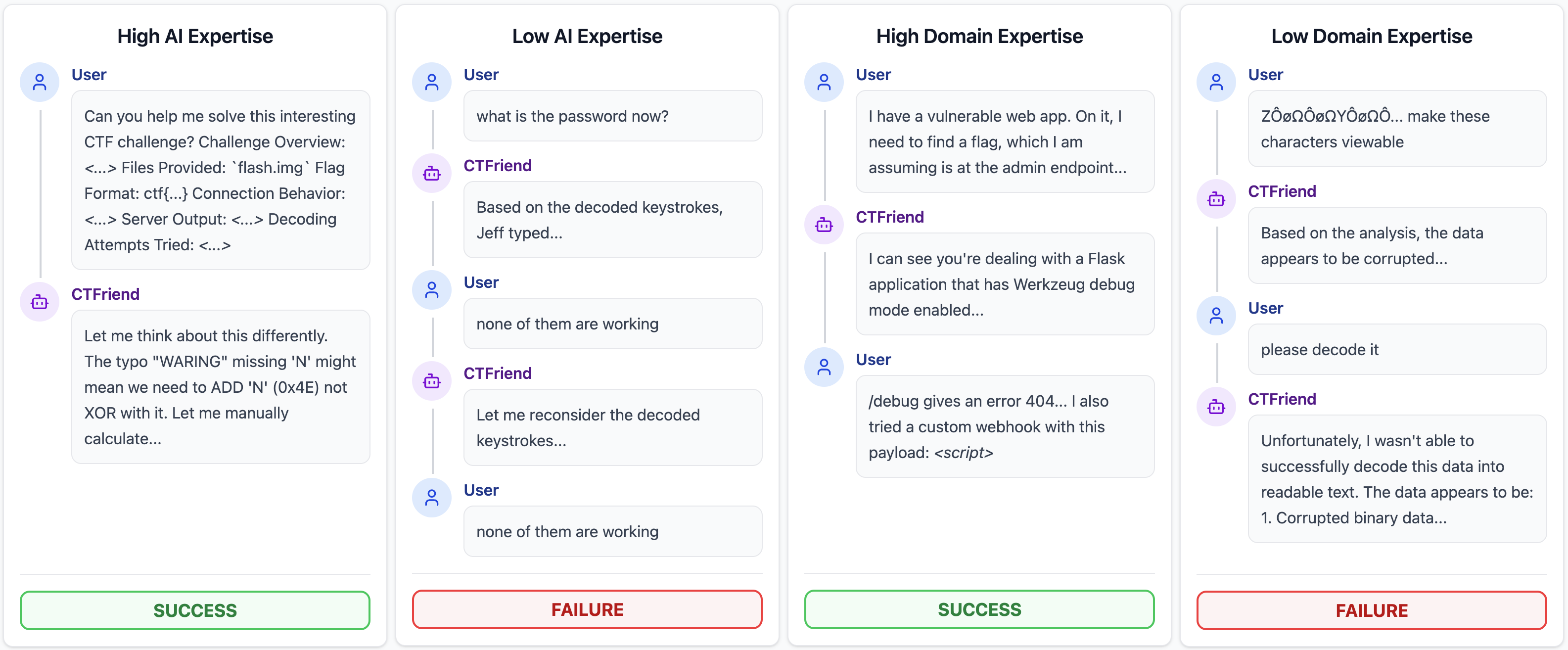}
  \caption{Representative Prompt Examples}
  \label{fig: prompt_example}
\end{figure*}

\noindent$\bullet$\textit{ Prompting Strategy:} Users who reported greater AI expertise more frequently used prompt-engineering strategies (66\% prevalence) and more often achieved successful outcomes in their episodes (\texttt{SucR} = 62\%) vs low-expertise users who saw 23\% and 27\% respectively. The most common strategy was \textit{chain-of-thought}, where users collaborated with AI to explore an challenge over multiple logical steps. For reference, representative examples of high- and low-expertise prompts and subsequent agent responses are shown in Fig \ref{fig: prompt_example}. Additionally, participants who reported zero or novice-level AI expertise also saw higher prevalence of AI errors (42\%) compared to experts (17\%).  This pattern held across all error categories, but especially for guardrail errors, 91\% of which occurred in novice AI users' logs.  This indicates that much of the frustration reported by participants in \ref{sec:rq1} may not have been due to technical limitations of the model, but user error. For example, a participant asking \textit{"what is <target>'s ssn?"} led to guardrail activation and AI task refusal.

\finding{Users with high AI expertise employed effective prompt engineering strategies and saw higher success rates while novices used low-quality prompts and experienced more errors and task failures.}

\noindent$\bullet$\textit{  Prompting Rabbit-Holes:} In the pre-CTF survey, most participants reported preferring to fully understand AI-provided code or system commands before running them. In practice, however, participants, including those that reported high AI and CTF expertise, typically executed suggested commands or code immediately then replied with the resulting system output. We observe this behavior in (i) short latencies between assistant command/code suggestions and the user’s subsequent system-output message (median $\Delta t=$ 23s), and (ii) the extremely low incidence (<1\%) of users rejecting AI-suggested commands after review (\textit{User: Command/Code Rejection}).  From their manual analysis, the CTF author suggested that this manifested as the user essentially serving as an interface between the AI and the system without appearing to assess problem-solving steps critically for themselves.  They specifically identified the AI's tendency to suggest next steps as a critical feature of this behavioral loop.  The low median time gap ($\Delta t=$14s) between the AI suggesting next steps (\textit{AI: Unprompted, Provide Next Steps}) and the user approving or executing them (\textit{User: Simple Approval}) may be indicative of this pattern.  On difficult challenge tasks, the expert assessed that users with low CTF expertise became trapped in "rabbit holes" where, after providing insufficient context in an initial task prompt, they engaged in iteration on an invalid solution thread -- the same failure pattern the expert had observed in autonomous agents logs \ref{sec:agents}. Additionally, the expert assessed that when users appeared to lack the necessary domain knowledge to proceed in the challenge, interaction quality would degrade. They highlighted examples of participants repeatedly asking the assistant to explain content or making unproductive requests (e.g., \textit{``give the flag''}). According to the expert, once this pattern emerged, the assistant’s suggestions became less actionable, reducing the chance of breaking the pattern.  

\noindent$\bullet$\textit{  Domain Knowledge-Guided Prompting:} 
\label{sec:domain_knowledge}
As such, the CTF author suggests two complementary value-adds presented by high domain expertise: first, expert users articulated their understanding of the challenge, including hypotheses, candidate directions, and intermediate observations, which helped steer the assistant toward a plausible approach. Second, they provided execution context that the model would not otherwise have, such as selected relevant tool outputs, error messages, and environment details.
For example, a high-scoring team contributed substantive domain signals (e.g., \textit{"there appears to be intentional delay in server ping response"}) and incorporated intermediate results into subsequent prompts, whereas a lower-scoring team provided substantially less context alongside command or script output.
\finding{Users with low CTF expertise were sucked into "rabbit holes" on harder challenges, while high CTF and AI expertise players achieved higher scores due to their ability to inject domain knowledge into the agent's context.}




\paragraph{Risks and Rewards} In the course of our analysis we observed AI's potential to both cause harm and benefit, depending on how it is applied.

\noindent$\bullet$\textit{ The Agentic Slot Machine:} As discussed in Finding $\mathcal{F}_6$, the most common strategy across all players was to delegate the full challenge to the agent.  However, an interesting pattern emerged in how that strategy was applied.  Specifically, among the lowest-scoring and lowest domain-knowledge teams, it was common to repeatedly delegate a full challenge (\texttt{Del2} = 23\%). In effect, this is rolling the dice that the LLM's temperature could lead to random generation of a correct solution.  Upon manual inspection of these instances, we observed that this pattern was distinctly unstructured: instead of managing a set of distinct contexts, users would offer no instructions outside the challenge description and keep the prompt chain in the same context window - negating the benefits of \textit{self-consistency}.  Rather than intentional strategy, this appears to be an emergent example of \textit{variable-reward reinforcement}, more commonly known as the \textit{"slot machine effect"}~\cite{choliz2010experimental}.  We infer this because the behavior saw a lower success rate (\texttt{Del2Fail} = 13\%) than the overall success rate for novice users (27\%), and thus served no strategic advantage.  From a risk-reward perspective, however, it is possible that the chance of achieving success with little effort expended per attempt made this pattern of \textit{``answer shopping''} more appealing to these participants despite the low chance of success. This would align with previous work regarding AI use in time-sensitive tasks, such as quizzes \cite{zheng2025aieducation}.  Additional research has also investigated the addictive nature of this interaction and the tendency for users to become invested in the potential low-effort reward at the expense of more optimal strategies \cite{shen2025aiaddiction}.

\finding{Participants with lower levels of domain and AI expertise would engage in "answer shopping", chain-regenerating outputs in hopes of a desirable solution, likely due to its low-cost and high potential reward.}

\noindent$\bullet$\textit{ An Uplifting Tool:}  Furthermore, zero-experience and novice participants who engaged deeply with the agent and were more resilient to failure achieved higher scores than their peers.   Notably, rather than the assistant solving problems for them, these participants asked questions that allowed them to iterate through challenges.  For example, User 21 reported themselves as having no experience in any CTF domain, which was corroborated by coding, with 60\% of their tasks \textit{indicative of low knowledge}.  Nonetheless, they applied repeated information-seeking, then leveraged the knowledge gained into new tasks for the agent.  As such, they were able to complete multiple challenges, and finished with the second-highest score among zero-experience players.  In the post-CTF survey, User 21 expressed that their greatest limitation was not knowing what questions to ask.  Nonetheless, they reported that the agent helped them solve multiple challenges they would have not been able to otherwise.  This theme of AI providing a crutch to some participants and as a learning and performance aid to others is a finding that has been observed in other research concerning AI use in competitive, time-pressured environments \cite{zheng2025aieducation}.

\finding{Participants with zero or novice-level domain experience were able to use AI to as a learning tool and constructive competition aid if they engaged in high-quality prompting, especially information-seeking, and were more resilient in their problem-solving attempts.}

\section{RQ$_\textbf{3}$:  Agents vs. Human Teams} 
RQ$_2$ reveals that human-AI collaboration is often constrained by the human side of the loop, including ineffective prompting and gaps in domain knowledge.
\label{rq3}
%
%
%
In contrast, autonomous agents self-direct key parts of the workflow, including prompt construction, tool use, and sometimes even model selection.
At the same time, the models themselves have increased their internal knowledge banks, which likely include substantial security knowledge (e.g., Sonnet 4.5 is trained on security-related data~\cite{anthropic2025sonnet45}), and expanded their thinking patterns and input contexts.
This motivates RQ$_\textbf{3}$: how do autonomous CTF agents compare to human teams on the same challenge set, can they outperform humans, and what limitations remain?

%

%


\subsection{Methodology}

\vspace{2pt}\noindent\textbf{Benchmark construction.} 
We converted CTF challenges into a machine-readable format. Following the structured schema~\cite{shao2024nyuctfbench,zhang2025cybench}, we created one JSON specification per challenge containing the same information available to the human participants, including the challenge name, description, category, points, and a list of challenge artifacts. 
%
We will release our benchmark dataset later (due to anonymity) in \cite{projectWebsite}.

\vspace{2pt}\noindent\textbf{Agents evaluated.}
We evaluated four autonomous agents: a coding assistant (Claude Code), two CTF-focused solvers (the NYU agent~\cite{shao2024nyuctfbench}and Cybench~\cite{zhang2025cybench}), and a proprietary security assistant that was among the top-performing agents in the Neurogrid AI-only CTF~\cite{neurogridctf}.
Agent details are provided in Appendix~\S\ref{Evaluated_agents}.

\vspace{2pt}\noindent\textbf{Experimental protocol and model selection.}
Each agent was given up to \textit{three attempts} per challenge per model. An attempt terminated when the agent produced a valid flag or exhausted its budget. 
Models were drawn from the Claude-family (Sonnet-4.5, Opus-4.1, and Haiku-3.5), the same models that were available to human participants via \agentName{}. 

\vspace{2pt}\noindent\textbf{API cost budgets.}
We set per-challenge API cost limits to bound resource usage. 
For \textit{Sonnet-4.5}, the budgets were \texttt{\$3} (first attempt), \texttt{\$5} (second), and \texttt{\$10} (third). For \textit{Opus-4.1}, the budgets were \texttt{\$10}, \texttt{\$15}, and \texttt{\$20}. For \textit{Haiku-3.5}, the budgets were \texttt{\$1}, \texttt{\$3}, and \texttt{\$5}. These limits follow the common practice of using an approximately \texttt{\$3-per-challenge} budget in prior work~\cite{shao2025craken,zhang2025cybench}, and we scale budgets across models with slight adjustments to account for differences in per-token pricing (Opus is more expensive and Haiku is cheaper)~\cite{anthropic2026pricing}.

  \begin{figure*}[t]
  \centering
  \includegraphics[width=1.0\linewidth]{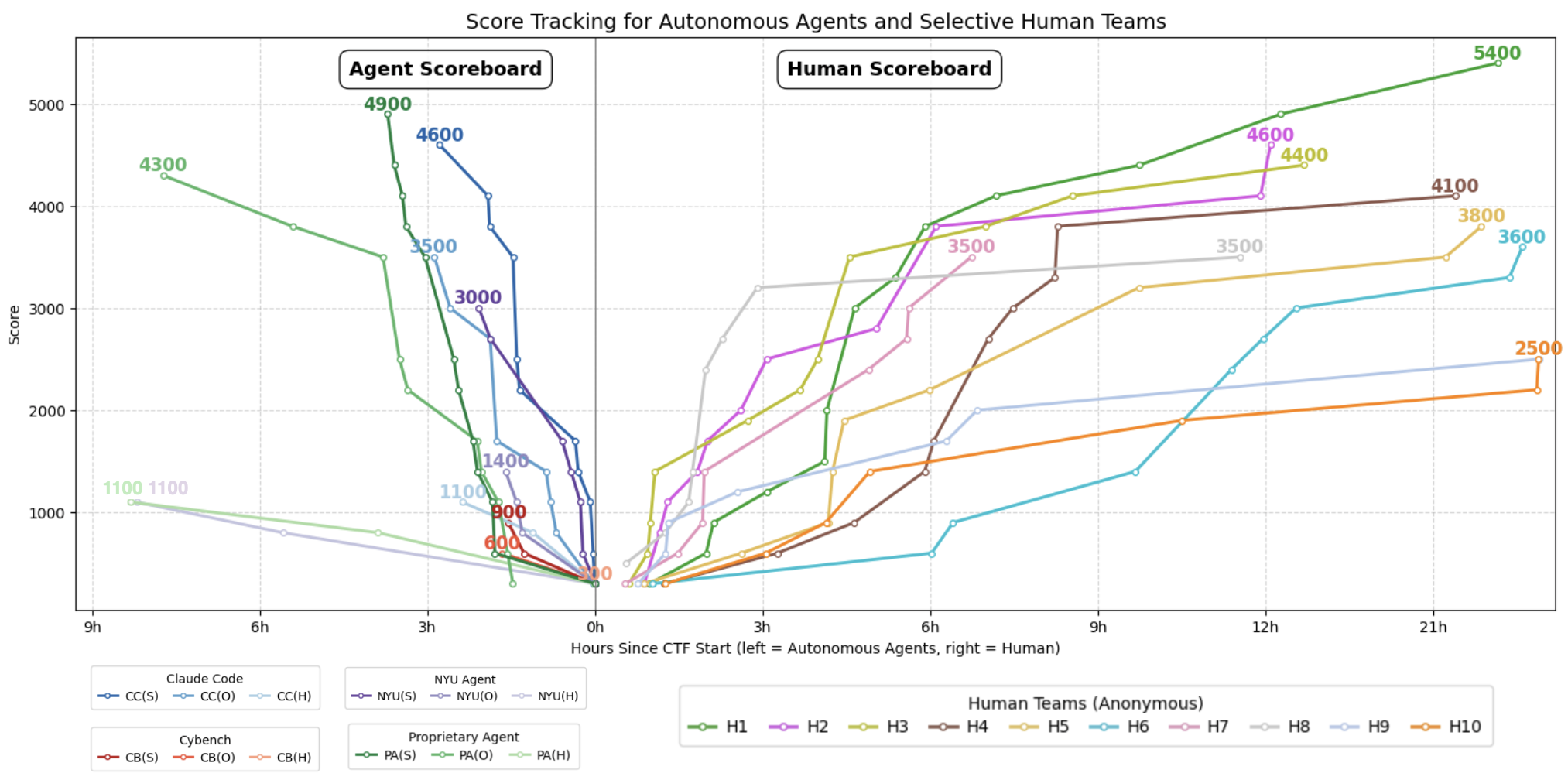}
  \caption{Score Tracking for Agent Teams and Top 10 Human Teams. (S), (O) and (H) mean Sonnet-4.5, Opus-4.1, and Haiku-3.5 models respectively.}
  \label{fig: score_tracking}
\end{figure*}

\begin{table*}[t]
\centering
\small
\setlength{\tabcolsep}{4.2pt}
\renewcommand{\arraystretch}{1.15}

\begin{adjustbox}{width=\linewidth}
\begin{tabular}{llcccccccccccccccc}
\toprule
\textbf{Model} & \textbf{Agent}
& \textbf{Overall Success Rate}
& \multicolumn{3}{c}{\textbf{Rev}}
& \multicolumn{3}{c}{\textbf{Crypto}}
& \multicolumn{3}{c}{\textbf{Forensics}}
& \multicolumn{3}{c}{\textbf{Web}}
& \multicolumn{3}{c}{\textbf{Other}} \\
\cmidrule(lr){4-6}\cmidrule(lr){7-9}\cmidrule(lr){10-12}\cmidrule(lr){13-15}\cmidrule(lr){16-18}
& & & \textbf{Succ.} & \textbf{Time} & \textbf{Tok.}
    & \textbf{Succ.} & \textbf{Time} & \textbf{Tok.}
    & \textbf{Succ.} & \textbf{Time} & \textbf{Tok.}
    & \textbf{Succ.} & \textbf{Time} & \textbf{Tok.}
    & \textbf{Succ.} & \textbf{Time} & \textbf{Tok.} \\
\midrule

Sonnet-4.5 & Proprietary Agent 
& 12/17 (70.59\%)
& 2/3 (66.67\%)     & 55.59 & 6.83
& 3/3 (100\%)     & 47.02 & 0.2
& 3/4 (75\%)   & 34.69 & 4.53
& 2/2 (100\%)     & 7.98 & 0.21
& 2/5 (40\%)    & 83.68 & 11.13 \\
        & Claude Code
& 11/17 (64.71\%)
& 2/3 (66.67\%) & 18.67 & 4.49
& 3/3 (100\%)   & 3.00  & 0.29
& 3/4 (75\%)    & 12.88 & 3.04
& 1/2 (50\%)    & 11.00 & 3.78
& 2/5 (40\%)    & 19.00 & 4.68 \\
          & NYU Agent
& 7/17 (41.18\%)
& 1/3 (33.33\%) & 34.74 & 9.26
& 3/3 (100\%)   & 12.10 & 3.16
& 1/4 (25\%)    & 32.00 & 5.98
& 0/2 (0\%)     & 10.58 & 2.19
& 2/5 (40\%)    & 28.16 & 6.35 \\
          & Cybench
& 3/17 (17.65\%)
& 0/3 (0\%)     & 12.04 & 2.73
& 0/3 (0\%)     & 21.83 & 2.37
& 2/4 (50\%)    & 10.22 & 1.69
& 0/2 (0\%)     & 11.97 & 1.45
& 1/5 (20\%)    & 21.11 & 1.73 \\
\addlinespace[0.35em]\midrule\addlinespace[0.35em]

Opus-4.1 & Proprietary Agent 
& 10/17 (58.82\%)
& 1/3 (33.33\%)     & 84.18 & 43.85
& 3/3 (100\%)     & 11.12 & 2.85
& 3/4 (75\%)   & 50.69 & 22.59
& 1/2 (50\%)     & 69.49 & 26.37
& 2/5 (40\%)    & 66.15 & 27.28 \\
        & Claude Code
& 8/17 (47.06\%)
& 0/3 (0\%)     & 21.33 & 27.97
& 2/3 (66.67\%) & 17.67 & 8.24
& 3/4 (75\%)    & 24.50 & 15.47
& 1/2 (50\%)    & 13.50 & 10.94
& 2/5 (40\%)    & 40.40 & 15.43 \\
        & NYU Agent
& 4/17 (23.53\%)
& 0/3 (0\%)     & 35.56 & 22.48
& 1/3 (33.33\%) & 43.36 & 30.56
& 1/4 (25\%)    & 38.78 & 26.23
& 0/2 (0\%)     & 27.30 & 17.38
& 2/5 (40\%)    & 38.97 & 18.08 \\
        & Cybench
& 2/17 (11.76\%)
& 0/3 (0\%)     & 15.46 & 6.94
& 0/3 (0\%)     & 30.25 & 10.12
& 1/4 (25\%)    & 15.21 & 6.85
& 0/2 (0\%)     & 26.39 & 7.03
& 1/5 (20\%)    & 20.08 & 6.19 \\
\addlinespace[0.35em]\midrule\addlinespace[0.35em]

Haiku-3.5 & Proprietary Agent 
& 3/17 (17.65\%)
& 0/3 (0\%)     & 233.23 & 9.00
& 1/3 (33.33\%)     & 150.43 & 6.08
& 1/4 (25\%))   & 216.94 & 6.76
& 0/2 (0\%)     & 203.91 & 9.00
& 1/5 (20\%)    & 202.62 & 7.61 \\

        & Claude Code
& 3/17 (17.65\%)
& 0/3 (0\%)     & 12.33  & 0.54
& 1/3 (33.33\%) & 9.67   & 0.58
& 1/4 (25\%)    & 11.10  & 0.54
& 0/2 (0\%)     & 19.50  & 0.21
& 1/5 (20\%)    & 36.80  & 0.67 \\
         & NYU Agent
& 3/17 (17.65\%)
& 0/3 (0\%)     & 144.53 & 3.60
& 1/3 (33.33\%) & 136.01 & 2.24
& 1/4 (25\%)    & 58.15  & 1.29
& 0/2 (0\%)     & 77.52  & 1.75
& 1/5 (20\%)    & 110.67 & 2.23 \\
         & Cybench
& 1/17 (5.88\%)
& 0/3 (0\%)     & 14.35  & 0.52
& 0/3 (0\%)     & 47.49  & 0.39
& 1/4 (25\%)    & 12.08  & 0.27
& 0/2 (0\%)     & 9.21   & 0.17
& 0/5 (0\%)     & 32.90  & 0.39 \\
\bottomrule
\end{tabular}
\end{adjustbox}

\caption{Performance and cost comparison across models and \textbf{challenge categories}. Succ.\ refers to Success rate, representing what percentage of challenges the agent solved; Time represents the average time spent for one challenge in minutes; Tok.\ denotes the average token usage for one challenge in dollars.}
\label{tab: performance_byCategory}
\end{table*}

\subsection{Agents beat most humans}
\label{sec:agents}

Figure~\ref{fig: score_tracking} compares 12 fully autonomous agent teams (four frameworks $\times$ three models) against the top-10 human teams.
For agents, \textit{``Hours Since CTF Start''} is computed as cumulative runtime by sequentially adding (i) the time to the first successful solve for each challenge, or (ii) the full time budget when the agent fails (i.e., exhausts its budget).
The strongest AI agent team is the proprietary agent with Sonnet-4.5, which reaches \texttt{4900} points (second place on the human scoreboard) in roughly four hours of cumulative runtime, at a cost of \$95.32 in API usage.
The next best runs are Claude Code with Sonnet-4.5 (\texttt{4600}) and the proprietary agent with Opus-4.1 (\texttt{4300}), which would rank third and fifth, respectively, among the top-10 human teams.
A detailed per-agent analysis of performance and cost is in Appendix \S~\ref{Agent team result analysis}.

\finding{An advanced agent outperforms most human teams while requiring only $\sim$1/5 of the cumulative runtime.}

From the results and the agent logs, we observe that:

\noindent$\bullet$\textit{ Agent design matters when paired with strong models } 
Across Sonnet-4.5 and Opus-4.1, switching from an assistant-style framework (Proprietary Agent and Claude Code) to a CTF-specific agent framework (NYU agent and Cybench) results in a significant drop in performance.
The assistant-style frameworks focus on longer-horizon planning and, most importantly, provide flexible tool interaction.
Agents are not restricted to a small subset of custom APIs; they can install tools on demand and execute commands directly in the terminal.
This flexibility allows the agent to adapt to unexpected issues during a challenge.
In contrast, the lower-performing CTF-specific agents (NYU agent and Cybench) rely on custom tool wrappers (e.g., \texttt{read\_file} instead of \texttt{cat}) that the model may be less familiar with, or on fixed tool sets that cannot adapt to unexpected issues (e.g., a non-interactive network connection tool for an interactive challenge).
\finding{Long-horizon planning and flexible, interactive tool support are the two framework features most associated with top-tier autonomous agent performance.}

\noindent$\bullet$\textit{ Model capability is the dominant bottleneck.}
Across frameworks, Sonnet-4.5 yields the highest success rate across four agents on average (48.53\%), followed by Opus-4.1 (35.29\%), while Haiku-3.5 collapses performance (14.71\%).
Even the proprietary agent drops sharply from \texttt{4900} (Sonnet-4.5) to \texttt{1100} (Haiku-3.5), and other agents converge to similarly low scores under Haiku (e.g., \texttt{1100} for NYU and \texttt{300} for Cybench).
\finding{Model capability ultimately sets the ceiling. Even strong tooling and architecture cannot compensate for a weak base model, and under weaker models the different frameworks largely converge to similarly low performance.
}

\subsubsection{What makes a challenge hard for the AI?} 
Challenges that are hard for human teams are not necessarily hard for autonomous agents (Table \ref{tab:human_ai_difficulty_difference}), and the reverse is also true.
In general, agents performed better than humans across all categories, except \texttt{rev}, where humans showed a higher solved rate. Specifically, for example, on a \texttt{crypto challenge} (C2), the solved rate for humans was 51.16\% and for agents was 75\% ( \textcolor{green!60!black}{$\Delta 23.84\%\uparrow$}). On a \texttt{logic-level challenge} (O3), 44.19\% human teams solved it, while about 92\% agents solved (\textcolor{green!60!black}{$\Delta 51.16\%\uparrow$}).

When looking deeper, we observed that agents performed best on challenges that minimized task complexity and environmental interactions. In essence, challenge easy for the agents were ones that could be reduced to ``write code and run it'' (e.g crypto challenges). Where agents struggled, especially the CTF specific agents were challenges that required numerous, complex tooling workflows and stateful, multi-step analysis and interactions (e.g. web challenges). Table~\ref{tab: performance_byCategory} breaks down agent teams' performance by challenge category, including success rate and resource usage.

\noindent$\bullet$\textit{ Reverse Engineering (Rev):}
Rev is among the hardest categories for agents and often incurs the highest time and token costs.
Only the strongest Sonnet-4.5 configurations can solve more than a small subset of Rev challenges, and overall agent solved rate is \textcolor{red!70!black}{$\Delta 10.47\%\downarrow$} lower than humans.
This gap arises because Rev tasks frequently require complex tool workflows and multi-step reasoning where agents remain brittle (see~\S\ref{The power of pair hacking}).


\noindent$\bullet$\textit{ Cryptography (Crypto):}
Crypto is the most agent-friendly category.
Four agent teams (Proprietary Agent and Claude Code with Sonnet-4.5, NYU with Sonnet-4.5, and the Proprietary Agent with Opus-4.1) achieve a 100\% solved rate with relatively low time and token cost, while humans are \textcolor{red!70!black}{$\Delta 10.47\%\downarrow$} lower.
Crypto is comparatively easy for agents because these challenges only involve basic file reading for the initial analysis, followed by writing and running a solver script. LLMs were precisely designed for these tasks~\cite{joel2024survey}.

\noindent$\bullet$\textit{ Forensics:}
Agents substantially outperform humans on forensics challenges (\textcolor{green!60!black}{$\Delta 10.47\%\uparrow$} in solved rate).
Many tasks in this category involve extracting and transforming evidence from artifacts (e.g., corrupted files) using straightforward, repeatable workflows that involve mostly simple tool sequences. Once the agent creates an analysis workflow, it can iterate much more quickly than a human.


\noindent$\bullet$\textit{ Web:}
Web challenges are difficult for both the autonomous agents and humans. These challenges require an understanding of network protocol exploitation, packet analysis, and involve intensive environment interaction. The proprietary agent, due to its strong interactive tool support, was able to leverage the capabilities of Sonnet-4.5 to solve both web challenges.
With respect to the the NYU and Cybench agents, they failed to solve these challenges even with the strongest model, largely due to limited support for interactive web sessions (e.g., maintaining state, handling multi-step workflows, and reacting to dynamic responses).


\noindent$\bullet$\textit{ Other:}
This represents the five challenges across OSINT, coding, and hardware.
Agents show moderate and relatively consistent performance (roughly 20\%-40\% success rate).
These tasks are often decomposable into structured subtasks (search, transform, implement), although physical hardware constraints can limit automation.
Overall, agent teams outperform humans by \textcolor{green!60!black}{$\Delta 55.62\%\uparrow$} points in solved rate.


\finding{ Crypto and forensics are comparatively easier for the agent teams than for humans, whereas reverse engineering remains harder for agents and favors human expertise; these mismatches suggest that human and AI strengths are complementary.}

\subsection{The power of pair hacking}
\label{The power of pair hacking}

Agents excel at high-throughput exploration and tedious work~\cite{Pentest_Testing_Corp}, but they are brittle when execution depends on environment interaction or when a wrong hypothesis leads to repeated retries. These failure modes are often less challenging from a human perspective. Humans are strong at inferring intent, prioritizing promising directions, and recognizing dead ends.
This complementarity makes \textit{pair hacking} a natural winner: a human-in-the-loop workflow in which the agent does the heavy lifting while a human provides sparse, high-leverage guidance.



In our study, we discovered that most agent failures were due to:
\noindent\textit{(\ding{202}) Problem solving loops} Agents can get stuck in unproductive retry cycles, repeatedly exploring the same incorrect path. For example, with Haiku-3.5, we often observed a ``reset'' behavior where the agent re-read the prompt and artifacts and restarted reconnaissance, effectively discarding prior progress.
\noindent\textit{(\ding{203}) Tooling/Environment limitations} Agents sometimes fail because the required interaction is not supported by the tool layer or execution environment. When this happens, they can spend the remaining budget trying to debug an issue that is outside their control. For example, the NYU agent failed to solve web challenges because the network tool was not designed for interactive web sessions. 


The proprietary agent we studied supports both fully autonomous runs and a collaborative \textit{pair hacking} mode. The developers built a UI that allows humans to monitor the agent’s trajectory and intervene at critical decision points. Interventions can be lightweight, such as providing a short hint, or operational, such as modifying the execution environment by installing additional tools or performing an interaction on the agent’s behalf. We asked the developers to revisit the five challenges their agent failed to solve autonomously and attempt them in pair hacking mode. Using this workflow, they solved two additional challenges, increasing the final score to \textit{5700} and placing first on both the human and agent leaderboards. 

\noindent\textbf{Case study.}
One of the two additional solves enabled by pair hacking was a reverse engineering challenge involving a GUI binary that revealed the flag after a correct password was entered.
The autonomous agent attempted to run the binary to observe its behavior, but the sandbox was command-line only (headless), causing execution to fail due to the missing display.
The agent then exhausted its budget trying to ``fix'' execution rather than progressing on the underlying Rev task.
In pair hacking mode, a human with reverse-engineering experience provided three lightweight interventions that redirected the trajectory: they (i) broke the execution-fix loop and shifted the agent to static analysis, (ii) resolved an external dependency by manually supplying a blocked dictionary file (\texttt{rockyou.txt}) after a firewall prevented download, and (iii) prompted broader code inspection after verifying the decryption logic, which led the agent to discover and replicate thousands of \texttt{random()} calls required to recover the correct flag. This case illustrates how sparse human guidance can bypass environment friction, interrupt unproductive loops, and restore forward progress without replacing the agent’s core problem-solving work.


\finding{This pair hacking scenario highlights the advantage of human-in-the-loop workflows: the agent can run autonomously while humans provide sparse steering and verification.}

\section{Threats to Validity}
\label{sec:validity}

\paragraph{Internal Validity}
While we manually cross-referenced individual scores to reduce missatribution (\S~\ref{sec:CTF_performance}), some participant submissions may reflect flags derived from team effort, potentially overstating individual performance.  Additionally, we elected to have hybrid coding performed by a single, trained, expert coder due to the technical nature of the corpus and codes (Section \ref{sec:quali_methodology}).  This introduces the threat of idiosyncratic coding, which we mitigated by adopting a two-stage exploration-confirmation codebook development process, however some risk remains. Furthermore, although we assessed single-coder stability via a blinded test–retest procedure, intra-coder agreement measures consistency rather than correctness; a degree of residual construct validity risk remains for ambiguous cases.


\paragraph{External Validity} As our study combines survey responses with qualitative content analysis of AI chat logs within a specific context, we do not claim statistical generalizability. Instead, our goal is \textbf{analytic} generalization/transferability: we identify recurring patterns and mechanisms that may plausibly apply to similar users, tasks, and deployment contexts, supported by detailed reporting of study context and participant/task characteristics and the full coding system (Appendix \S~\ref{sec:codebook}).   Furthermore, our coding process is sound, but the nature of the corpus is such that there was variance in the number of logs obtained from each participant, and the number of tasks contained in each log.  Although we normalize by episode/turn counts, residual differences in task complexity may still affect code prevalence (\S~\ref{sec:competition_behavior}). Finally, because LLM behavior varies across model versions, interfaces, and tool configurations, observed patterns may shift over time and may not directly generalize beyond the specific system configuration studied.

\section{Discussion and Conclusion}

In this section, we distill our findings into four key discussion themes that conclude with practical takeaways for the security community. 

\subsection{For CTF Organizers}

CTFs have historically co-evolved with automation. 
%
Our results observe that a similar transition is now underway for AI agents: in our setting, a fully autonomous agent nearly matched the performance of the top human teams ($\mathcal{F}_{11}$).
If a CTF's goal is to evaluate security reasoning under pressure, organizers should anticipate that autonomous agents can now solve a non-trivial fraction of standard challenge archetypes, and quickly.
A naïve countermeasure would be to add inordinately difficult or convoluted challenges to confuse AI; however, such challenges would also discourage human players, defeating the educational agenda that motivates CTFs. 
One design adaptation could be to include tasks that are solvable by humans but are \textit{operationally hard} for current agents because they require robust interaction with complex, stateful tooling or environments.
In our experiments, agents commonly failed from brittle execution and interaction, such as reproducing a challenge in a specific environment, driving a browser workflow with multiple steps, synchronizing with services that require waiting or state transitions, or handling tasks where progress depends on careful, iterative inspection rather than a single-shot script.
Such challenges will remain fair to humans if organizers provide clear setup instructions, stable artifacts, and good debugging signals, while resisting fully autonomous ``pipeline'' solving ($\mathcal{F}_{12}$-$\mathcal{F}_{15}$).

\begin{takeaway}{Takeaway 1}%
CTFs will need to adapt to the use of AI agents and deploy interactive challenges that engage the human players while being operationally hard for AI agents.
\end{takeaway}%

\subsection{For Cybersecurity Educators}
When used as a scaffold rather than a substitute for reasoning, AI assistance can lower the entry barrier for CTF novices ($\mathcal{F}_{10}$), improve skill acquisition, and help novices progress more quickly toward independent problem solving.
This was observed especially in participants who prioritized information-seeking and were resilient to failure.  
However, unscaffolded learning can potentially lead to performance without transfer: students complete tasks more efficiently in the presence of AI guidance, yet may fail to develop skills that persist when guidance is removed. Zhou et al. illustrate this risk in an introductory programming setting: LLM-generated hints improved immediate debugging success, but the advantages disappeared when AI support was withdrawn \cite{zhou2025aieducation}.  Essentially, if an LLM routinely proposes the next action, learners may bypass the metacognitive work of deciding what to do next and why ($\mathcal{F}_{9}$), weakening strategic planning, error diagnosis, and critical evaluation of evidence. In cybersecurity education, where expertise depends on recognizing patterns across unfamiliar systems and reasoning under uncertainty, such reliance is especially problematic. 

AI might be turned into a meaningful benefit, however, as new research argues for theory-driven, adaptive scaffolding that is contingent on learners’ demonstrated understanding and that fades as competence increases \cite{cohn2026aieducation}.  For example, a future CTF could include an ``educational'' division with an AI assistant that could: (i) require learners to articulate their reasoning (e.g., prompting for hypotheses and justification before revealing hints), (ii) provide tiered assistance that starts with conceptual guidance rather than executable steps, and (iii) progressively reduce specificity as the learner demonstrates improvement.  Such an approach could preserve the motivational and accessibility benefits of timely help while mitigating cognitive offloading.  

\begin{takeaway}{Takeaway 2}%
Our findings demonstrate an opportunity for security educators to adapt existing work in AI scaffolding to the context of solving security challenges, in order to retain the educational value of CTFs and security challenges in general while helping students learn to leverage AI constructively.
\end{takeaway}%

\subsection{For Emerging Security Practitioners}
Our study suggests that alongside traditional domain expertise, \textit{AI literacy} is becoming an increasingly important skill for security practitioners.
As we observed, in the AI-assisted CTF setting, the highest-performing teams were not simply those with the strongest CTF background, but those who could maximize AI capability under time pressure ($\mathcal{F}_{4}$).
Consistent with our interaction analysis, participants with strong AI-use skills (e.g., high-quality prompting and structured iteration) were often able to compensate for gaps in CTF domain knowledge and still achieve strong outcomes, whereas some domain-expert participants who struggled to elicit actionable guidance or validate outputs saw their performance degrade despite their underlying expertise ($\mathcal{F}_{5}$).

\begin{takeaway}{Takeaway 3}
Our findings show that ``knowing security'' and ``using AI effectively'' are now complementary competencies rather than substitutes. It is in the interest of next-generation practitioners to train explicitly for AI-enabled workflows by learning how to scope tasks, provide grounded context, and verify outputs, using AI to accelerate exploration without relinquishing human responsibility for correctness and risk.    
\end{takeaway}

\subsection{For AI Agent Developers}

Our results show that the most effective workflow is often neither ``AI assistance'' nor full autonomy, but a \textit{human-in-the-loop} pattern in which the agent performs high-throughput work while a human provides sparse, high-leverage supervision ($\mathcal{F}_{8}$-$\mathcal{F}_{10}$).
Autonomous agents excel at rapid reconnaissance, broad exploration, and repetitive tasks, yet we repeatedly observed brittle failure modes that lightweight human intervention can resolve ($\mathcal{F}_{15}$).
For example, agents enter unproductive loops, issuing repetitive commands or attempting  brute-force methods even if evidence suggests that the current path is unlikely to succeed.
Practically, this motivates the injection of a Human-in-the-loop layer in autonomous agentic workflow design for security tasks, which (1) allows humans to monitor progress and re-plan when a strategy stops paying off; (2) expose intermediate evidence and assumptions so humans can quickly verify or correct them; and (3) seamlessly integrate both high-level or precise feedback in the agentic loop.
%
However, building such a human-in-the-loop layer would require us to rethink the balance of autonomy and human intervention in the design of agentic systems, in order to involve the human in a manner that is usable, while also simultaneously maximizing agentic autonomy.

\begin{takeaway}{Takeaway 4}
At the task of solving CTF challenges, agentic systems work best when there is a human in the loop to help the agent strategize, identify conclusive evidence, and recover from deadlocks.  This paradigm is human-AI collaboration in its truest form, and exposes a valuable opportunity for effective use of AI in security that may go beyond CTFs. 
\end{takeaway}

\cleardoublepage
\appendix
\section*{Ethical Considerations}
This study was reviewed and approved by our Institutional Review Board (IRB).
All participants provided informed consent prior to data collection.
We minimized risk by collecting only data necessary for the study (e.g., survey responses and \agentName{} interaction logs), avoiding sensitive personal identifiers, and limiting access to the raw data to the research team.
We report results in aggregate and anonymize any quoted excerpts to prevent re-identification.


\section*{Open Science}
To support reproducibility, we release study artifacts at \url{https://anonymous.4open.science/w/HumanAI_Collaboration-12E5/}, including (1) the pre- and post-survey instruments, (2) the qualitative codebook, (3) the \agentName{} code implementation, and (4) anonymized experiment logs. We will release the benchmark dataset after the paper is published due to anonymity considerations.



{
\small
\bibliographystyle{ieeetr}
\bibliography{main}

@inproceedings{ji2025measuring,
  title={Measuring and Augmenting Large Language Models for Solving Capture-the-Flag Challenges},
  author={Ji, Zimo and Wu, Daoyuan and Jiang, Wenyuan and Ma, Pingchuan and Li, Zongjie and Wang, Shuai},
  booktitle={CCS},
  year={2025}
}

@inproceedings{shao2024nyuctfbench,
     author = {Shao, Minghao and Jancheska, Sofija and Udeshi, Meet and Dolan-Gavitt, Brendan and xi, haoran and Milner, Kimberly and Chen, Boyuan and Yin, Max and Garg, Siddharth and Krishnamurthy, Prashanth and Khorrami, Farshad and Karri, Ramesh and Shafique, Muhammad},
     booktitle = {Advances in Neural Information Processing Systems},
     pages = {57472--57498},
     title = {NYU CTF Bench: A Scalable Open-Source Benchmark Dataset for Evaluating LLMs in Offensive Security},
     url = {https://proceedings.neurips.cc/paper_files/paper/2024/file/69d97a6493fbf016fff0a751f253ad18-Paper-Datasets_and_Benchmarks_Track.pdf},
     volume = {37},
     year = {2024}
}

@inproceedings{zhang2025cybench,
title
=
{Cybench: A Framework for Evaluating Cybersecurity Capabilities and Risks of Language Models},
author
=
{Andy K Zhang and Neil Perry and Riya Dulepet and Joey Ji and Celeste Menders and Justin W Lin and Eliot Jones and Gashon Hussein and Samantha Liu and Donovan Julian Jasper and Pura Peetathawatchai and Ari Glenn and Vikram Sivashankar and Daniel Zamoshchin and Leo Glikbarg and Derek Askaryar and Haoxiang Yang and Aolin Zhang and Rishi Alluri and Nathan Tran and Rinnara Sangpisit and Kenny O Oseleononmen and Dan Boneh and Daniel E. Ho and Percy Liang},
booktitle
=
{The Thirteenth International Conference on Learning Representations},
year
=
{2025},
url
=
{https://openreview.net/forum?id=tc90LV0yRL},
}

@inproceedings{
  abramovich2025enigma,
  title={En{IGMA}: Interactive Tools Substantially Assist {LM} Agents in Finding Security Vulnerabilities},
  author={Talor Abramovich and Meet Udeshi and Minghao Shao and Kilian Lieret and Haoran Xi and Kimberly Milner and Sofija Jancheska and John Yang and Carlos E Jimenez and Farshad Khorrami and Prashanth Krishnamurthy and Brendan Dolan-Gavitt and Muhammad Shafique and Karthik R Narasimhan and Ramesh Karri and Ofir Press},
  booktitle={Forty-second International Conference on Machine Learning},
  year={2025},
  url={https://openreview.net/forum?id=Of3wZhVv1R}
}

@article{shao2025craken,
  title={CRAKEN: Cybersecurity LLM Agent with Knowledge-Based Execution},
  author={Shao, Minghao and Xi, Haoran and Rani, Nanda and Udeshi, Meet and Putrevu, Venkata Sai Charan and Milner, Kimberly and Dolan-Gavitt, Brendan and Shukla, Sandeep Kumar and Krishnamurthy, Prashanth and Khorrami, Farshad and Karri, Ramesh and Shafique, Muhammad},
  journal={arXiv preprint arXiv:2505.17107},
  year={2025},
  url={https://arxiv.org/abs/2505.17107}
}

@article{zhuo2025training,
  title={Training Language Model Agents to Find Vulnerabilities with CTF-Dojo},
  author={Zhuo, Terry Yue and Wang, Dingmin and Ding, Hantian and Kumar, Varun and Wang, Zijian},
  journal={arXiv preprint arXiv:2508.18370},
  year={2025}
}

@article{buckley2021regulating,
  title     = {Regulating Artificial Intelligence in Finance: Putting the Human in the Loop},
  author    = {Buckley, Ross P. and Zetzsche, Dirk Andreas and Arner, Douglas W. and Tang, Brian},
  journal   = {Sydney Law Journal},
  volume    = {43},
  year      = {2021},
  note      = {University of Hong Kong Faculty of Law Research Paper 2021/016, UNSW Law Research Paper No. 23-29},
  url       = {https://ssrn.com/abstract=3831758}
}

@misc{DARPA,
  title        = {Artificial Intelligence Cyber Challenge {(AIxCC)}},
  author       = {DARPA},
  howpublished = {\url{https://aicyberchallenge.com/}},
  year         = {2025}
}

@article{song2015darpa,
  title={The darpa cyber grand challenge: A competitor's perspective},
  author={Song, Jia and Alves-Foss, Jim},
  journal={IEEE Security \& Privacy},
  volume={13},
  number={6},
  pages={72--76},
  year={2015},
  publisher={IEEE}
}

@inproceedings{yang2023language,
  title={Language agents as hackers: Evaluating cybersecurity skills with capture the flag},
  author={Yang, John and Prabhakar, Akshara and Yao, Shunyu and Pei, Kexin and Narasimhan, Karthik R},
  booktitle={Multi-Agent Security Workshop@ NeurIPS'23},
  year={2023}
}

@article{zheng2025aieducation,
    author = {Jiayu Zheng and Lingxin Hao and Kelun Lu and Ashi Garg and Mike Reese and Melo-Jean Yap and I-Jeng Wang and Xingyun Wu and Wenrui Huang and Jenna Hoffman and Ariane Kelly and My Le and Ryan Zhang and Yanyu Lin and Muhammad Faayez and Anqi Liu},
    title = {Do Students Rely on AI ? Analysis of Student-ChatGPT Conversations from a Field Study},
    journal = {arXiv preprint arXiv:2508.20244v1},
    year = {2025}
}

@inproceedings{zhou2025aieducation,
title = {Impact of LLM Feedback on Learner Persistence in Programming},
author = {Zhou, Yiqiu and Pankiewicz, Maciej and Paquette, Luc and Baker, Ryan},
booktitle={Proceedings of the 33rd International Conference on Computers in Education},
year = {2025},
month = {12},
pages = {},
}

@article{cohn2026aieducation,
    author = {Clayton Cohn and Siyuan Guo and Surya Rayala and Hanchen David Wang and Naveeduddin Mohammed and Umesh Timalsina and Shruti Jain and Angela Eeds and Menton Deweese and Pamela J. Osborn Popp Rebekah Stanton and Shakeera Walker and Meiyi Ma and Gautam Biswas},
    title = {Evidence-Decision-Feedback: Theory-Driven Adaptive Scaffolding for LLM Agents},
    journal = {arXiv preprint arXiv:2508.20244v1},
    year = {2025}
}

@inproceedings{shen2025aiaddiction,
author = {Shen, M. Karen and Yoon, Dongwook},
title = {The Dark Addiction Patterns of Current AI Chatbot Interfaces},
year = {2025},
isbn = {9798400713958},
publisher = {Association for Computing Machinery},
address = {New York, NY, USA},
url = {https://doi.org/10.1145/3706599.3720003},
doi = {10.1145/3706599.3720003},
booktitle = {Proceedings of the Extended Abstracts of the CHI Conference on Human Factors in Computing Systems},
articleno = {514},
numpages = {7},
location = {
},
series = {CHI EA '25}
}

@misc{neurogridctf,
  title        = {{Neurogrid CTF:} The ultimate AI security showdown},
  author       = {Hack The Box},
  howpublished = {\url{https://ctf.hackthebox.com/event/details/neurogrid-ctf-the-ultimate-ai-security-showdown-2712}},
  year         = {2025}
}

@article{spracklenpackage,
  title={Package Hallucinations: How LLMs Can Invent Vulnerabilities},
  author={Spracklen, Joseph and Jadliwala, Murtuza}
}

@misc{picoctf2026,
  author        = {{Carnegie Mellon University}},
  title        = {picoCTF - CMU Cybersecurity Competition},
  year         = {2026},
  howpublished = {\url{https://picoctf.org/}}
}

@misc{csawctf2025,
  author        = {{NYU OSIRIS Lab}},
  title        = {CSAW Capture the Flag (CTF)},
  year         = {2025},
  howpublished          = {\url{https://www.csaw.io/ctf}}
}

@misc{defconctf,
  author = {{DEF CON Communications, Inc.}},
  title = {DEF CON® Hacking Conference - Capture the Flag Archive},
  howpublished = {\url{https://defcon.org/html/links/dc-ctf.html}},
  year = {2026}
}

@misc{anthropic2026pricing,
  author = {{Anthropic}},
  title = {Claude API Pricing},
  year = {2026},
  howpublished = {\url{https://platform.claude.com/docs/en/about-claude/pricing}}
}

@misc{anthropic2025sonnet45,
  author = {{Anthropic}},
  title = {Introducing Claude Sonnet 4.5},
  month = {September},
  year = {2025},
  howpublished = {\url{https://www.anthropic.com/news/claude-sonnet-4-5}}
}

@inproceedings{chatgpt_vulnerable_manage_2024,
author = {Liu, Peiyu and Liu, Junming and Fu, Lirong and Lu, Kangjie and Xia, Yifan and Zhang, Xuhong and Chen, Wenzhi and Weng, Haiqin and Ji, Shouling and Wang, Wenhai},
title = {Exploring ChatGPT's capabilities on vulnerability management},
year = {2024},
isbn = {978-1-939133-44-1},
publisher = {USENIX Association},
address = {USA},
booktitle = {Proceedings of the 33rd USENIX Conference on Security Symposium},
articleno = {46},
numpages = {18},
location = {Philadelphia, PA, USA},
series = {SEC '24}
}

@inproceedings{huynh2025detecting,
  title={Detecting code vulnerabilities using llms},
  author={Huynh, Larry and Zhang, Yinghao and Jayasundera, Djimon and Jeon, Woojin and Kim, Hyoungshick and Bi, Tingting and Hong, Jin B},
  booktitle={2025 55th Annual IEEE/IFIP International Conference on Dependable Systems and Networks (DSN)},
  pages={401--414},
  year={2025},
  organization={IEEE}
}

@article{nong2024chain,
  title={Chain-of-thought prompting of large language models for discovering and fixing software vulnerabilities},
  author={Nong, Yu and Aldeen, Mohammed and Cheng, Long and Hu, Hongxin and Chen, Feng and Cai, Haipeng},
  journal={arXiv preprint arXiv:2402.17230},
  year={2024}
}

@article{yu2024insight,
  title={An Insight into Security Code Review with LLMs: Capabilities, Obstacles and Influential Factors},
  author={Yu, Jiaxin and Liang, Peng and Fu, Yujia and Tahir, Amjed and Shahin, Mojtaba and Wang, Chong and Cai, Yangxiao},
  journal={arXiv preprint arXiv:2401.16310},
  year={2024}
}

@inproceedings{kim2025logs,
  title={Logs In, Patches Out: Automated Vulnerability Repair via $\{$Tree-of-Thought$\}$$\{$LLM$\}$ Analysis},
  author={Kim, Youngjoon and Shin, Sunguk and Kim, Hyoungshick and Yoon, Jiwon},
  booktitle={34th USENIX Security Symposium (USENIX Security 25)},
  pages={4401--4419},
  year={2025}
}

@article{zhang2023well,
  title={How well does LLM generate security tests?},
  author={Zhang, Ying and Song, Wenjia and Ji, Zhengjie and Meng, Na and others},
  journal={arXiv preprint arXiv:2310.00710},
  year={2023}
}

@article{wu2023deceptprompt,
  title={Deceptprompt: Exploiting llm-driven code generation via adversarial natural language instructions},
  author={Wu, Fangzhou and Liu, Xiaogeng and Xiao, Chaowei},
  journal={arXiv preprint arXiv:2312.04730},
  year={2023}
}

@article{huang2023catastrophic,
  title={Catastrophic jailbreak of open-source llms via exploiting generation},
  author={Huang, Yangsibo and Gupta, Samyak and Xia, Mengzhou and Li, Kai and Chen, Danqi},
  journal={arXiv preprint arXiv:2310.06987},
  year={2023}
}

@article{fang2024llm,
  title={Llm agents can autonomously exploit one-day vulnerabilities},
  author={Fang, Richard and Bindu, Rohan and Gupta, Akul and Kang, Daniel},
  journal={arXiv preprint arXiv:2404.08144},
  year={2024}
}

@article{lin2025llm,
  title={LLM-HyPZ: Hardware Vulnerability Discovery using an LLM-Assisted Hybrid Platform for Zero-Shot Knowledge Extraction and Refinement},
  author={Lin, Yu-Zheng and Ghimire, Sujan and Nandimandalam, Abhiram and Camacho, Jonah Michael and Tripathi, Unnati and Macwan, Rony and Shao, Sicong and Rafatirad, Setareh and Yasaei, Rozhin and Satam, Pratik and others},
  journal={arXiv preprint arXiv:2509.00647},
  year={2025}
}

@article{li2025vulsolver,
  title={VULSOLVER: Vulnerability Detection via LLM-Driven Constraint Solving},
  author={Li, Xiang and Su, Yueci and Liu, Jiahao and Lin, Zhiwei and Hou, Yuebing and Gao, Peiming and Zhang, Yuanchao},
  journal={arXiv preprint arXiv:2509.00882},
  year={2025}
}

@article{weissberg2025llm,
  title={LLM-based Vulnerability Discovery through the Lens of Code Metrics},
  author={Weissberg, Felix and Pirch, Lukas and Imgrund, Erik and M{\"o}ller, Jonas and Eisenhofer, Thorsten and Rieck, Konrad},
  journal={arXiv preprint arXiv:2509.19117},
  year={2025}
}

@article{sheng2024lprotector,
  title={Lprotector: An llm-driven vulnerability detection system},
  author={Sheng, Ze and Wu, Fenghua and Zuo, Xiangwu and Li, Chao and Qiao, Yuxin and Hang, Lei},
  journal={arXiv preprint arXiv:2411.06493},
  year={2024}
}

@inproceedings{chen2025exp,
  title={Exp-Arch: A Novel LLM-Powered Approach for Facilitating Exploit Primitive Assessment in the Linux Kernel},
  author={Chen, Zuxin and Li, Zhi and Song, Zhanwei and Shi, Zhiqiang and Sun, Limin},
  booktitle={2025 IEEE 31th International Conference on Parallel and Distributed Systems (ICPADS)},
  pages={01--10},
  year={2025},
  organization={IEEE}
}

@article{yu2024llm,
  title={Llm-enhanced software patch localization},
  author={Yu, Jinhong and Chen, Yi and Tang, Di and Liu, Xiaozhong and Wang, XiaoFeng and Wu, Chen and Tang, Haixu},
  journal={arXiv preprint arXiv:2409.06816},
  year={2024}
}

@inproceedings{kulsum2024case,
  title={A case study of llm for automated vulnerability repair: Assessing impact of reasoning and patch validation feedback},
  author={Kulsum, Ummay and Zhu, Haotian and Xu, Bowen and d'Amorim, Marcelo},
  booktitle={Proceedings of the 1st ACM International Conference on AI-Powered Software},
  pages={103--111},
  year={2024}
}

@article{xu2025revisiting,
  title={Revisiting Vulnerability Patch Localization: An Empirical Study and LLM-Based Solution},
  author={Xu, Haoran and Zhi, Chen and Han, Junxiao and Zhao, Xinkui and Yin, Jianwei and Deng, Shuiguang},
  journal={arXiv preprint arXiv:2509.15777},
  year={2025}
}

@article{zhang2025patch,
  title={PATCH: Empowering Large Language Model with Programmer-Intent Guidance and Collaborative-Behavior Simulation for Automatic Bug Fixing},
  author={Zhang, Yuwei and Jin, Zhi and Xing, Ying and Li, Ge and Liu, Fang and Zhu, Jiaxin and Dou, Wensheng and Wei, Jun},
  journal={ACM Transactions on Software Engineering and Methodology},
  year={2025},
  publisher={ACM New York, NY}
}

@misc{projectWebsite,
  author        = {{Anonmous authors}},
  title        = {Poject Website},
  year         = {2026},
  howpublished          = {\url{https://anonymous.4open.science/w/CTF_website-8A4B/}}
}

@misc{maxqda,
  title        = {The top 1 qualitative data analysis software with the best AI integration},
  year         = {2026},
  howpublished          = {\url{https://www.maxqda.com/}}
}

@article{choliz2010experimental,
  title={Experimental analysis of the game in pathological gamblers: Effect of the immediacy of the reward in slot machines},
  author={Ch{\'o}liz, Mariano},
  journal={Journal of Gambling Studies},
  volume={26},
  number={2},
  pages={249--256},
  year={2010},
  publisher={Springer}
}

@misc{Pentest_Testing_Corp,
  title        = {5 Hard Truths About AI Pentest Agents vs Humans},
  year         = {2026},
  howpublished          = {\url{https://meetcyber.net/5-hard-truths-about-ai-pentest-agents-vs-humans-0f6e4b05a816}}
}

@article{joel2024survey,
  title={A survey on llm-based code generation for low-resource and domain-specific programming languages},
  author={Joel, Sathvik and Wu, Jie and Fard, Fatemeh},
  journal={ACM Transactions on Software Engineering and Methodology},
  year={2024},
  publisher={ACM New York, NY}
}

@article{zou2026ctfagent,
  title={CTFAgent: An LLM-powered Agent for CTF Challenge Solving},
  author={Zou, Yuwen and Liu, Jia and Fan, Wenjun},
  journal={Journal of Information Security and Applications},
  volume={96},
  pages={104305},
  year={2026},
  publisher={Elsevier}
}

@inproceedings{mousavi2024investigation,
  title={An investigation into misuse of java security apis by large language models},
  author={Mousavi, Zahra and Islam, Chadni and Moore, Kristen and Abuadbba, Alsharif and Babar, M Ali},
  booktitle={Proceedings of the 19th ACM Asia Conference on Computer and Communications Security},
  pages={1299--1315},
  year={2024}
}

@inproceedings{abramovichenigma,
  title={EnIGMA: Interactive Tools Substantially Assist LM Agents in Finding Security Vulnerabilities},
  author={Abramovich, Talor and Udeshi, Meet and Shao, Minghao and Lieret, Kilian and Xi, Haoran and Milner, Kimberly and Jancheska, Sofija and Yang, John and Jimenez, Carlos E and Khorrami, Farshad and others},
  booktitle={Forty-second International Conference on Machine Learning}
}

@article{mayoral2025cybersecurity,
  title={Cybersecurity AI: The World's Top AI Agent for Security Capture-the-Flag (CTF)},
  author={Mayoral-Vilches, V{\'\i}ctor and Navarrete-Lozano, Luis Javier and Balassone, Francesco and Sanz-G{\'o}mez, Mar{\'\i}a and Chavez, Crist{\'o}bal RJ and de Torres, Maite del Mundo and Turiel, Vanesa},
  journal={arXiv preprint arXiv:2512.02654},
  year={2025}
}

@article{sun2024ai,
  title={AI hallucination: towards a comprehensive classification of distorted information in artificial intelligence-generated content},
  author={Sun, Yujie and Sheng, Dongfang and Zhou, Zihan and Wu, Yifei},
  journal={Humanities and Social Sciences Communications},
  volume={11},
  number={1},
  pages={1--14},
  year={2024},
  publisher={Palgrave}
}

@article{debnath2025comprehensive,
  title={A comprehensive survey of prompt engineering techniques in large language models},
  author={Debnath, Tonmoy and Siddiky, Md Nurul Absar and Rahman, Muhammad Enayetur and Das, Prosenjit and Guha, Antu Kumar and Rahman, Muhammad Rezaur and Kabir, HM},
  journal={TechRxiv},
  year={2025}
}
}

\appendix

\appendix

\section{Survey Design}
\label{Survey Design}
\vspace{4pt}\noindent\textbf{Pre-survey design} 
The pre-survey (17 questions) covered: (i) CTF background (experience, typical solve volume, self-rated skills by category), (ii) AI familiarity (prior use and comfort with prompting/LLM tools), (iii) expectations and trust (expected performance impact and trust in outputs), (iv) anticipated interaction patterns with AI, and (v) self-reported validation practices. 


\vspace{4pt}\noindent\textbf{Post-survey design}
The post-survey measured: (i) perceived helpfulness (including categories helped most/least and time savings), (ii) perceived output quality (e.g., clarity, completeness, actionability, correctness), (iii) failure experiences and coping strategies (responses to incorrect or unhelpful outputs), and (iv) future adoption and updated expectations. 

\vspace{4pt}\noindent\textbf{Construct Validity} While we designed our survey to minimize bias or participant misunderstanding (Section \ref{sec:methodology}, there still exists the threat of ambiguous constructs or inconsistent interpretation (e.g. participants may interpret “expertise,” “confidence,” “usefulness,” “hallucination,” or “trust” differently).  Additionally, the length of the post-CTF survey, and with many participants taking it directly following a long competition may have led to acquiescence or satisficing.  Further threats to construct validity, such as poorly calibrated self-assessment (e.g. over-estimation of expertise), impression management (e.g. overreporting desirable behaviors such as carefulness, skepticism) may be present, but represent part of the value of this study through their comparison to obvservable behavior during the competition (Section \ref{sec:competition_behavior}).

\section{The Design and Implementation of \agentName{}}
\label{sec:implementation}
To study human–AI collaboration in live CTF competitions, we design and deploy \agentName{}, an AI assistant tailored for CTF problem solving. \agentName{} is not intended to function as an automatic CTF solver; instead, it provides a controlled and observable collaboration platform that supports CTF participants in using AI assistant during competitions, while enabling researchers to observe and analyze human–AI interaction behaviors systematically. We released the platform in \cite{projectWebsite}.

  \begin{figure}[htbp]
    \centering
    \includegraphics[width=1.0\linewidth]{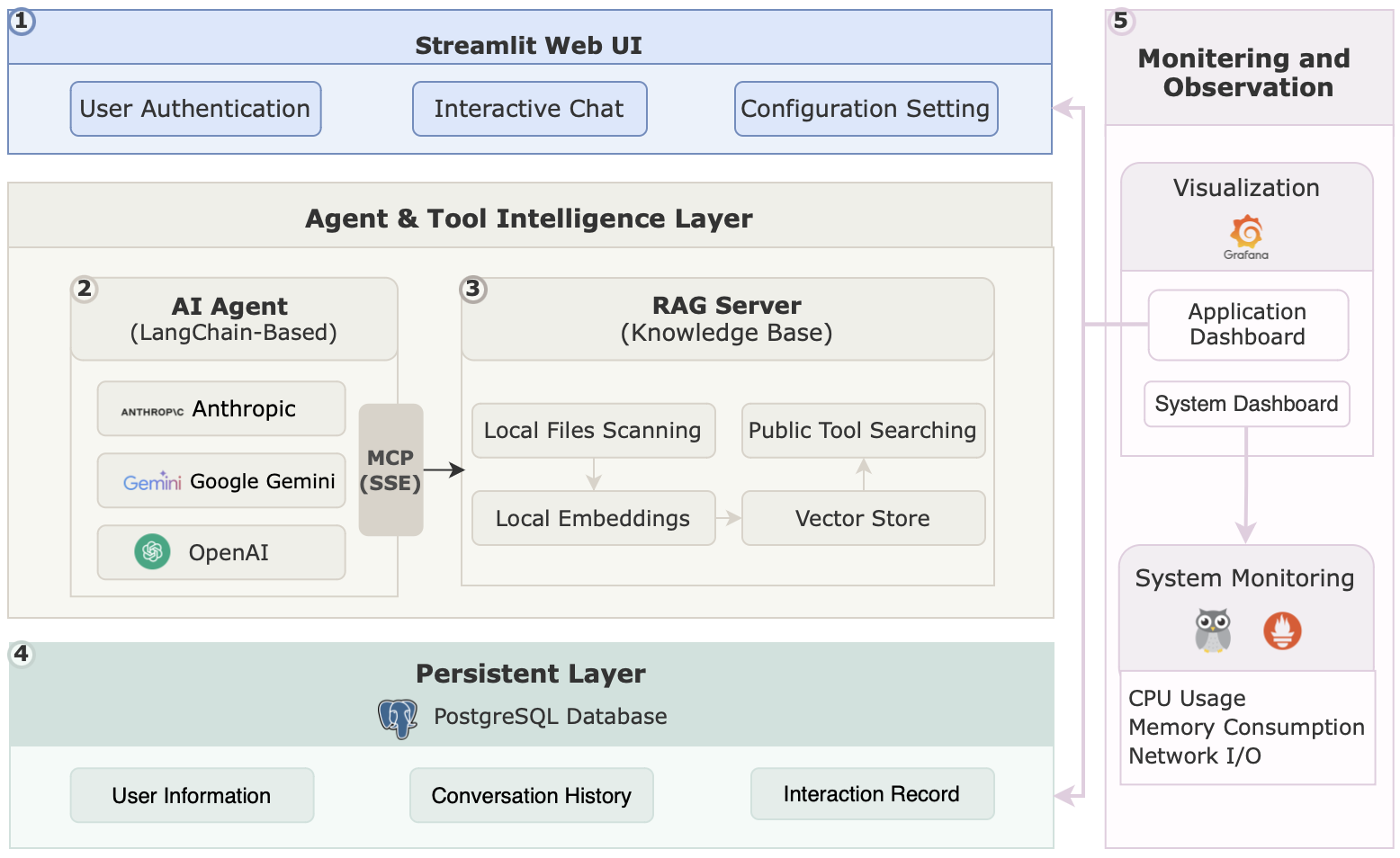}
    \caption{Overview of CTFriend }
    \label{fig:agent_workflow}
  \end{figure}
 
\subsection{Design Goals}

\agentName{} is designed to support the empirical study of human–AI collaboration in live CTF competitions, guided by the following design goals:

\noindent $\bullet$ \textit{Participant-Centered Collaborative Usability.} \agentName{} should adopt a conversational interface mirroring commonly used AI assistants, enabling natural and familiar interaction while preserving human decision-making authority. To support effective use in time-constrained CTF competitions, the system should minimize setup overhead and provide a unified interface for accessing multiple LLMs, reducing cognitive burden during live events.

\noindent $\bullet$ \textit{CTF-Specific Domain Orientation.} The system is explicitly tailored for CTF problem solving and should provide effective assistance for common CTF tasks and reasoning patterns,  improving the relevance of AI assistance in competitive CTF settings.

\noindent $\bullet$ \textit{Comprehensive Observability.} The platform should enable the detailed observation of human–AI interactions while also providing real-time monitoring of system health and application usage, ensuring reliable operation and high-quality data collection during live deployments.

\subsection{Implementation}

\agentName{} is implemented as a modular, tool-augmented conversational agent centered around a unified reasoning core. The agent is designed to integrate LLMs, external tool services, and persistent state management within the framework.

  \noindent $\bullet$ \textit{User Interface.} User interaction is handled through a lightweight Streamlit-based frontend, which renders the chat interface, displays conversation history, and captures user input and feedback. To handle conversation history and manage user sessions, a token-based authentication mechanism is used along with a uniquely identifiable token provided to each user. Successful authentication of the user's token results in the establishment of an authenticated session. This session would then handle all interactions and subsequent interactions for the duration of the session. Upon interaction with the frontend, all user input is passed via an API call to the corresponding backend system, resulting in an AI-generated response. Furthermore, runtime configuration parameters, such as LLM provider and model selection, can be selected via the Streamlit-based frontend and are later transmitted to the backend as structured metadata embedded in the API request.

  \noindent $\bullet$ \textit{Agent Core.} Using the LangChain framework, the responsibility for language model orchestration, conversational context management, and tool invocation falls to the backend agent core system. Support for multiple LLM providers (Gemini, OpenAI, Anthropic) is realized through dynamic client instantiation based on runtime configuration. Additionally, short-term conversational context is maintained in memory and backed to persistent storage, allowing for multi-turn reasoning and persistent conversation history and context, isolated for each conversation.
  
  \noindent $\bullet$ \textit{Knowledge Augmentation via Microservices.} A central aspect of CTFriend assistant is tool-augmented reasoning through the Model Context Protocol (MCP). The agent establishes persistent Server-Sent Events (SSE) connections to tool servers and dynamically registers available tools at startup. During inference, the agent is capable of autonomously invoking tools through structured requests, integrating returned results into its reasoning process, keeping the tool-using logic external to the agent.
  
  A retrieval-augmented generation (RAG) knowledge base is implemented as a standalone MCP tool server, constituting the primary knowledge augmentation mechanism in the system. The RAG service preprocesses local PDF and Markdown documents into semantically coherent splits, embeds them using a locally hosted sentence transformer model, and indexes the resulting vectors in an in-memory Facebook AI Similarity Search (FAISS) store. When invoked, the service performs a rapid and efficient semantic search, returning all relevant documents to the agent, grounding responses in authoritative local knowledge.
  
  \noindent $\bullet$ \textit{Management and Monitoring} All long-term interaction data, including user identities, conversation sessions, messages, and feedback signals, is persisted in a PostgreSQL database. By externalizing these data, the agent core remains stateless across executions, enabling reliable recovery. 
  
  System behavior and health are continuously observed through an integrated monitoring stack based on cAdvisor and Prometheus.  Additionally, Grafana provides visibility into container-level resource usage and application-level interaction patterns.

\begin{figure}[t]
  \centering
  \includegraphics[width=1.0\linewidth]{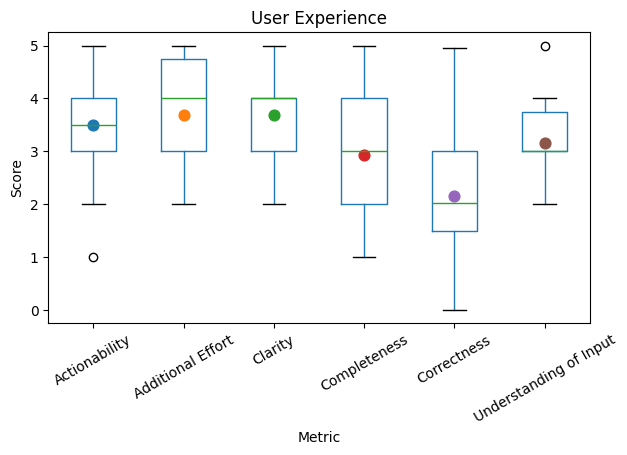}
  \caption{User Experience: user ratings across six evaluation dimensions. Box plots show the median and interquartile range. Open circles represent outliers, and colored dots indicate mean values.} 
  \label{fig: user_experience}
\end{figure}

\section{Qualitative Analysis}
\label{sec:quali_analysis}
\subsection{Coding Protocol}

\noindent $\bullet$ \textit{Deductive Codes: } AI Error codes were derived from Sun et al. \cite{sun2024ai}'s classification of LLM errors. 
While this work is primarily focused on issues regarding communication and social sciences, we believe it to be, with minor adaptations, a high-quality and comprehensive classification of AI errors encountered by participants. Prompt Engineering strategy codes development guided by the following Debnath et al.'s survey of LLM prompting techniques, adapted for user-agent interaction in a CTF environment \cite{debnath2025comprehensive}. 

Finally, base interaction pattern codes were derived from those developed for the pre- and post-CTF surveys, discussed in \ref{sec:methodology}.

\noindent $\bullet$ \textit{Deductive Codes: } Unless otherwise indicated, codes are horizontally non-exclusive and multiple codes may be applied to the same coding unit.  All units were coded and counted at the lowest subcode.  We provide a definition and frequency-only codebook in this appendix, while the full code system with per-code protocols, decision rules, instructions, and examples can be found at the paper website \cite{projectWebsite}.



\subsection{Supplemental Information}
\label{sec:qual_supplemental}

\noindent $\bullet$ \textit{Operationalized Variable Definitions: } 

\texttt{SucR}: \textit{Success} as a \% of \textit{Success}+\textit{Failure}.  \textit{Resolved but Unknown} excluded.  Essentially, "what \% of known outcome task episodes were \textit{Success} for the given demographic."

\texttt{Del2}: Binary Odds Ratio of \textit{Delegate Full Task} $\rightarrow$ \textit{Delegate Full Task} - \textit{Multiple Challenges in Same Context}.  Essentially, "what \% of \textit{Delegate Full Task} are followed by another \textit{Delegate Full Task} excluding those that co-occur with \textit{Multiple Challenges in Same Context} (i.e., the same challenge was regenerated).

\texttt{Del2Fail}: Binary Odds Ratio of \texttt{Del2Fail} $\rightarrow$ \textit{Failure}.  Essentially, "what \% of \texttt{Del2Fail} instances co-occured with a \textit{Failure} in the same task episode."

\begin{figure}[t]
  \centering
  \includegraphics[width=1.0\linewidth]{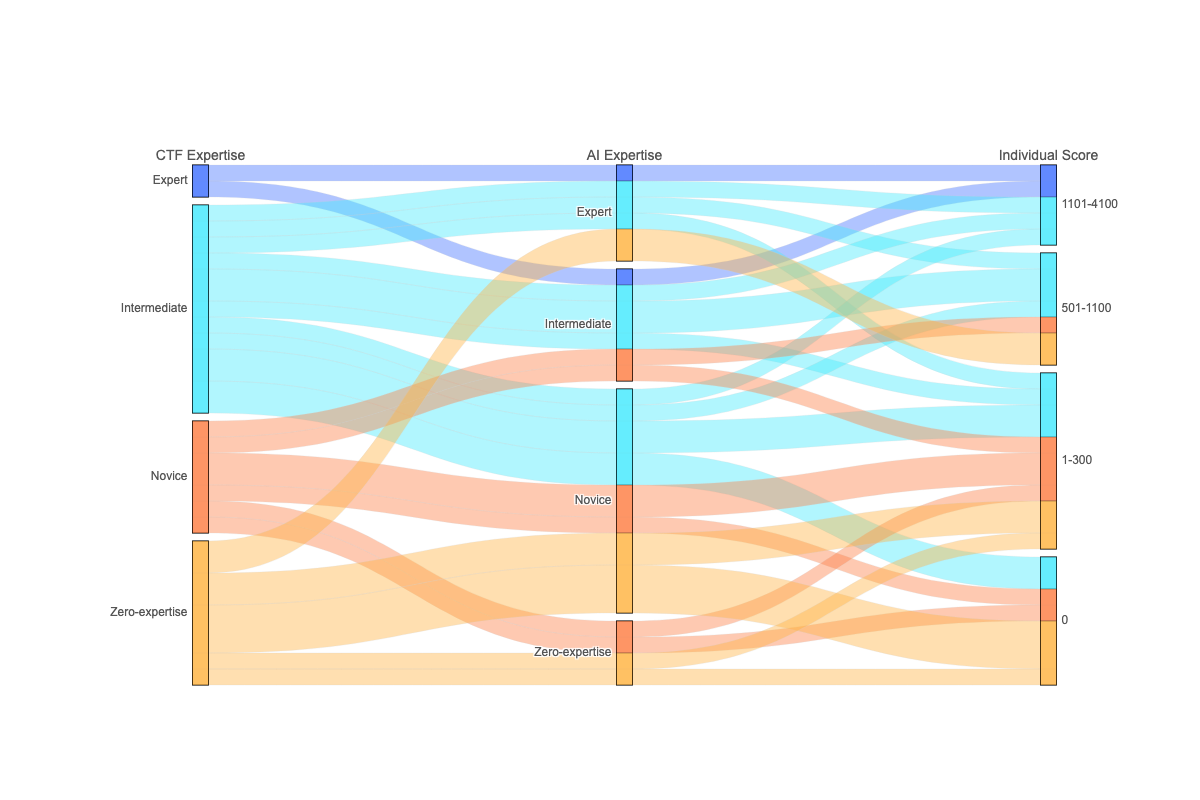}
  \caption{Correlation between Expertise and Score} 
  \label{fig: expertise vs. score}
\end{figure}

\noindent$\bullet$\textit{ CTF Scores:} We found that team scores were too granular as our analysis showed meaningful variation in behavior and performance within teams. Since the CTF platform tracked flag captures per player, we computed an individual score for each participant based on their recorded flag captures. To reduce misattribution, we manually cross-referenced each flag submission with user logs to confirm the submitting player worked on the challenge and that no other teammate completed it.  Fig. \ref{fig: expertise vs. score} illustrates the links between expertise and individual score.

\noindent$\bullet$\textit{ Interaction Length:}  Chat length varied wildly among participants, with some participants able to find a flag using only a few prompts, and others appearing to work on a problem over many iterations without finding a solution.  Overall, the average chat length was 13.6, with a median of 9, a minimum of 2 and a maximum of 56.  The average chat length is skewed by a small minority of participants who maintained the same conversation for most or all of the competition, attempting to solve multiple challenges within the same context window despite being instructed to not do so during onboarding. This behavior was observed exclusively among users who reported zero or novice-level AI experience and co-occurred strongly with low-quality prompts, suggesting that these users may not have had enough experience with AI agents to fully understand the researchers' instructions or general best practices for AI use.

\noindent$\bullet$\textit{ Changes  Over Time:} Over time, lower-scoring teams and less experienced AI users became more willing to delegate larger tasks to the AI, rather than asking pointed questions.  Specifically, prompts generally became larger in scope, allowing the AI more room for interpretation and generating strategies autonomously.  For example, early in the competition, a participant would be more likely to ask \textit{``What command should I use in <software> to achieve <specific output>?''}. Meanwhile, the same participant, in a later stage of the competition or facing a more difficult challenge would be more likely to ask \textit{``What should I try next?''}.  However, higher-scoring teams saw the opposite effect. These teams were more likely to attempt to delegate the full challenge to the model in the earliest stages of the competition, then increasingly break task into smaller subtasks before providing them to the agent at later stages.  This may to be a manifestation of a difference in strategic philosophy, with higher-scoring players prioritizing easy points by delegating the challenges the AI is most likely to complete, then increasing their level of effort and individual involvement as the competition wore on.

\section{Evaluated agents}
\label{Evaluated_agents}
\noindent$\bullet$\textit{ Claude Code.} We configured the Claude Code assistant as an autonomous agent for CTF challenge-solving. We design a specific prompt to encourage the agent to follow a traditional CTF solving workflow (i.e., interpret the prompt, reason about an approach, iteratively refine, and output a flag when found). 
To improve robustness, we prepared \textbf{six} prompt variants with minor adjustments for common failure scenarios (e.g., warning about decoy flags). All prompts are provided in Figure ~\ref{fig: claude code prompt}. We report results from the best-performing prompt configuration.

 \noindent$\bullet$\textit{ CTF Solving Agents} We used the NYU CTF automation framework~\cite{shao2024nyuctfbench} and the Cybench agent framework~\cite{zhang2025cybench}, which both provide agents specifically designed for CTF solving. We note the Cybench agent framework expects a different input format than the NYU CTF framework. Our benchmark uses the NYU CTF syntax so some benchmark conversion was necessary, but the information provided to the agent remains the same. It is given a challenge JSON and any files referenced in the JSON.



\noindent$\bullet$\textit{ Proprietary agent.} In November 2025, Hack The Box hosted the Neurogrid CTF, an AI-first CTF competition to benchmark AI agents capabilities. We contacted one of the top performing teams and asked them to evaluate their agent on our benchmark dataset. According to them, the agent was simply told to ``solve the challenge'' based on the provided challenge description and files.

\vspace{2pt}\noindent\textbf{Execution environment.} 
The proprietary agent was evaluated on a Ubuntu 22.04 server with a 64 core Intel(R) Xeon(R) CPU E5-4650 0 @ 2.70GHz and 756GB memory.
Claude Code and NYU agent were evaluated on a Debian (ARM 64-bit) virtual machine with an 8-core processor, 7.4 GB of memory, and a 124.8 GB virtual disk, running under a VirtualBox hypervisor with NAT networking.
Cybench agent was evaluated on a macOS 14.5 with a arm64 architecture, 8-core CPU, 16 GB of memory, and 460Gi of disk space.
For local challenges, the evaluation environment exposed the challenge artifacts (e.g., binaries, source files, or data files) to the agent through an accessible directory. For remote challenges, we hosted the target services on a separate server and allowed agents to connect to them remotely. This setup mirrors how participants interacted with local and remote challenges during the live event while keeping the evaluation procedure reproducible.


\section{Agent team result analysis}
\label{Agent team result analysis}
\noindent$\bullet$\textit{ Proprietary agent:} This agent achieves the first place on the agent scoreboard for every model variation and the second place on the human scoreboard with Sonnet-4.5 with a score of \texttt{4900}. This score represents a 70.59\% success rate. Looking at the proprietary agent configurations, there are small differences in solving performance between Sonnet-4.5 and Opus-4.1, but a large drop when using Haiku-3.5. We note a large gap in performance between the proprietary agent and the CTF specific agents. According to the developers, their agent is prompted to encourage long term planning and they do not provide the agent with custom tool wrappers, but rather allow it to run various tools interactively. This gives the agent flexibility as needed and resolves its own issues when they occur. Further, they believe that LLMs are more familiar with direct tool usages rather than custom function calls due to the baseline training data. 

\noindent$\bullet$\textit{ Claude Code:} Claude Code achieves the third place on the agent scoreboard for Sonnet-4.5 with a score of \texttt{4600}, sharing the same place with one human team. Claude Code also significantly outperforms agents built specifically for solving CTFs, except the proprietary agent. As with the proprietary agent, there is a dramatic drop in performance when using Haiku-3.5. With this model choice, all of the agents perform similarly poor. Therefore, while framework choice influences the agent's thinking patterns and tooling capabilities, model choice is still important in shaping the baseline reasoning capabilities.

\noindent$\bullet$\textit{ NYU Autonomous Framework:} The NYU Agent exhibits inconsistent performance across different models choices. As shown in Figure \ref{fig: score_tracking}, it trails behind both proprietary agent and Claude Code on Sonnet-4.5 and Opus-4.1, with \texttt{4300} points as the best. And it matches both agents on Haiku-3.5 (all for 1100 points). The jumps in performance across model choices suggests that the NYU Agent is dependent on advanced LLM reasoning capabilities to navigate its custom tooling layer. 

\noindent$\bullet$\textit{ Cybench LM Agent:} By contrast, Cybench ranked lowest across all model configurations, with scores of \texttt{900}, 600, and 300 points respectively. Its performance drop, particularly under weaker models, underscoring its limited capability of leveraging the advanced reasoning of the model. Moreover, the consistently low success rates (more details in Table \ref{tab: performance_byCategory}) indicate that Cybench may lack adaptive mechanisms to compensate for model limitations.



\subsection{Difficulty-based results}

\begin{table*}[t] 
\centering
\small
\setlength{\tabcolsep}{5.5pt}
\renewcommand{\arraystretch}{1.15}

\begin{adjustbox}{width=\linewidth}
\begin{tabular}{llcccccccccc}
\toprule
\textbf{Model} & \textbf{Agent}
& \textbf{Overall Success Rate}
& \multicolumn{3}{c}{\textbf{Easy Challenges}}
& \multicolumn{3}{c}{\textbf{Medium Challenges}}
& \multicolumn{3}{c}{\textbf{Hard Challenges}} \\
\cmidrule(lr){4-6} \cmidrule(lr){7-9} \cmidrule(lr){10-12}
& 
& 
& \textbf{Success Rate} & \textbf{Time} & \textbf{Token}
& \textbf{Success Rate} & \textbf{Time} & \textbf{Token}
& \textbf{Success Rate} & \textbf{Time} & \textbf{Token} \\
\midrule

Sonnet-4.5  & Proprietary Agent  & 12/17 (70.59\%) & 8/9 (88.89\%) & 32.78 & 2.35  & 3/6 (50.00\%)     &  69.85 & 6.33 & 1/2 (50.00\%)  & 83.43 & 9.16 \\
           & Claude Code & 11/17 (64.71\%) & 7/9 (77.78\%) & 16.05  & 2.44 & 3/6 (50.00\%) & 20.67  & 5.22 & 1/2 (50.00\%) & 22.00  & 6.69 \\
           & NYU Agent   & 7/17 (41.18\%) & 5/9 (55.56\%) & 14.97 & 3.04 & 1/6 (16.67\%) & 32.48 & 7.90 & 1/2 (50.00\%) & 33.64 & 11.78 \\
           & Cybench     & 3/17 (17.65\%) & 3/9 (33.33\%) & 13.10 & 2.16 & 0/6 (0\%)     & 15.35 & 1.97 & 0/2 (0\%)     & 31.82 & 1.17 \\

\addlinespace[0.4em]
\midrule
\addlinespace[0.4em]

Opus-4.1   & Proprietary Agent   & 10/17 (58.82\%) & 6/9 (66.67\%) & 48.46 & 20.19  & 3/6 (50.00\%)     &  65.88 & 23.57 & 1/2 (50.00\%)  & 69.9 & 25.50
           \\
          & Claude Code & 8/17 (47.06\%) & 5/9 (55.56\%) & 20.40 & 10.72 & 2/6 (33.33\%) & 33.50 & 15.47 & 1/2 (50.00\%) & 48.50 & 15.10 \\
           & NYU Agent   & 4/17 (23.53\%) & 3/9 (33.33\%) & 41.14 & 22.23 & 1/6 (16.67\%) & 21.99 & 10.37 & 0/2 (0\%)  & 52.00 & 41.42 \\
           & Cybench     & 2/17 (11.76\%) & 2/9 (22.22\%) & 14.15 & 5.74  & 0/6 (0\%)     & 24.34 & 10.85 & 0/2 (0\%)  & 29.90 & 6.68 \\

\addlinespace[0.4em]
\midrule
\addlinespace[0.4em]

Haiku-3.5  & Proprietary Agent  & 3/17 (17.65\%) & 2/9 (22.22\%) & 201.74 & 7.23  & 1/6 (16.67\%)     &  195.57 & 7.54 & 1/2 (50.00\%)  & 216.97 & 9.00\\
          & Claude Code & 3/17 (17.65\%) & 2/9 (22.22\%) & 13.80 & 0.44 & 1/6 (16.67\%) & 13.80 & 0.67 & 0/2 (0\%) & 63.00 & 0.65 \\
           & NYU Agent   & 3/17 (17.65\%) & 2/9 (22.22\%) & 111.31 & 1.69 & 1/6 (16.67\%) & 106.20 & 2.50 & 0/2 (0\%) & 71.93 & 2.83 \\
           & Cybench     & 1/17 (5.88\%) & 1/9 (11.11\%) & 17.44 & 0.30 & 0/6 (0\%)     & 32.00 & 0.47 & 0/2 (0\%) & 24.84 & 0.39 \\
           
\bottomrule
\end{tabular}
\end{adjustbox}

\caption{Performance comparison across models, agents, and \textbf{difficulty levels}. Time represents the average time spent for one challenge in minutes; Token denotes the average token usage for one challenge in dollars.}
\label{tab:performance_byDifficulty}
\end{table*} 

Table \ref{tab:performance_byDifficulty} illustrates the performance and cost of agents for challenges in difficulty levels. 

\noindent$\bullet$\textit{ Easy Challenges:} In the easy tier, proprietary agent with Sonnet-4.5 achieved the highest success rate (88.89\%), outperforming other agent-model combinations. It also maintained relatively low token cost (avg. 2.35 dollars per challenge). For Claude Code, although its success rate on this level was slightly lower than that of proprietary agent, it spent only a half time and very close token to solve these challenges, demonstrating both efficiency and effectiveness. The NYU Autonomous Framework under Sonnet showed decent performance (solved 5 out of 9 challenges) but consumed more tokens and time. For Cybench, despite the cost saving,it only solved one third challenges on easy challenges set, which is not ideal for a CTF-oriented agent. 
For Opus-4.1, success rate of all agents dropped to the same extent (by 22\%), while token usage significantly increased. Besides the factor of per-token pricing, it still shows the low time- and token-effectiveness. As mentioned in overall result \ref{tab:performance_byDifficulty}, Haiku-3.5 performed poorly across all agents, with all agents' success rate decreased to 22\%, except Cybench agent, suggesting that smaller models may offer cost benefits but at the expense of performance.

\noindent$\bullet$\textit{ Medium Challenges:} As challenge complexity increased, overall success rates declined, with both the proprietary agent and Claude Code (Sonnet-4.5) remaining the top success rate at 50\%, however, the proprietary agent required much more time on them. Other agents struggled, with NYU agent solving only one challenge (the same challenge) across all models and Cybench failing to solve any tasks. This suggests that larger models offer necessary reasoning capabilities for medium challenges, and well-designed agents like Claude Code can further enhance performance with modest time and cost overhead.

\noindent$\bullet$\textit{ Hard Challenges:} At the highest difficulty, a half agent and model combination achieved some success (solving one challenge), while all other combinations failed. Claude Code (Sonnet-4.5) maintained a competitive balance (with avg. 22 minutes and avg. 6.69 dollars per challenge), while the proprietary agent consistently asked for more token usage and longer time. Other agents underperformed significantly: NYU Framework could only see some progress when using Sonnet-4.5, and Cybench failed again. In summary, only the strong models and agents can handle difficult CTF challenges, and that inefficient model or agent limitations can lead to high cost without any benefit.


\begin{table}[t]
\centering
\scriptsize
\setlength{\tabcolsep}{7pt}
\renewcommand{\arraystretch}{1.15}
\begin{adjustbox}{width=\linewidth}
\begin{tabular}{lcccc}
\toprule
\textbf{Category} & \textbf{Challenge} & \textbf{Human} & \textbf{AI} & \textbf{$\Delta$ Performance} \\
\midrule

\multirow{4}{*}{Forensics}
 & F1 & 86.05\% & 100.00\% & \multirow{4}{*}{\textcolor{green!60!black}{$\Delta 44.76\%\uparrow$}} \\
 & F2 & 32.56\% &  41.47\% & \\
 & F3 & 11.63\% &  33.34\% & \\
 & F4 &  0.00\% &   0.00\% & \\
\midrule

\multirow{4}{*}{Cryptography}
 & C1 & 62.79\% & 33.33\% & \multirow{3}{*}{\textcolor{green!60!black}{$\Delta 10.47\%\uparrow$}} \\
 & C2 & 51.16\% & 75.00\% & \\
 & C3 & 25.58\% & 41.47\% & \\
\midrule

\multirow{4}{*}{Reverse Engineering}
 & R1 & 23.26\% & 25.00\% & \multirow{3}{*}{\textcolor{red!70!black}{$\Delta \textbf{10.47}\%\downarrow$}} \\
 & R2 & 34.88\% & 25.00\% & \\
 & R3 &  2.33\% &  0.00\% & \\
\midrule

\multirow{2}{*}{Web}
 & W1 &  9.30\% &  8.33\% & \multirow{2}{*}{\textcolor{green!60!black}{$\Delta 27.72\%\uparrow$}} \\
 & W2 &  4.65\% & 33.34\% & \\
\midrule

\multirow{5}{*}{Other}
 & O1 &  2.33\% &  0.00\% & \multirow{5}{*}{\textcolor{green!60!black}{$\Delta \textbf{55.62}\%\uparrow$}} \\
 & O2 & 39.53\% & 50.00\% & \\
 & O3 & 44.19\% & 91.67\% & \\
 & O4 &  0.00\% &  0.00\% & \\
 & O5 &  0.00\% &  0.00\% & \\
\bottomrule
\end{tabular}
\end{adjustbox}

\caption{Human vs.\ AI \textbf{solved rate} differences by \textbf{category}. Solved rate represent what percent of teams solved this challenge during on-site competition.}
\label{tab:human_ai_difficulty_difference}
\end{table}

\section{Agent failure reason analysis}

\subsubsection{Performance qualitative analysis:}

  
  

  \paragraph{Log-based Analysis.} Similar to that in RQ$_\textbf{2}$ \ref{sec:rq2}, we selected some autonomous log representatives to analyze their further failure reasons beyond their own architecture limitations.
  
  \noindent$\bullet$\textit{Log Selection:} In order to get a boarder reasons, we tried to seek a diverse set of samples. To keep consistency, we aligned selection criteria with those used in RQ$_\textbf{2}$ (\S~\ref{sec:rq2}), which means that we again selected one challenge from each of the three difficulty levels, identical to those chosen in RQ$_\textbf{2}$, with 9 representatives in total. 
  
  \noindent$\bullet$\textit{Manual Analysis:} To perform a qualitative analysis of the interactions and capabilities of the tested fully-automated AI teams, a series of 3 challenges was selected. These challenges were selected based on the following criteria: first, they must be representative of the 3 difficulty levels of created CTF challenges (Easy, Medium, and Hard). Second, they must all be challenges failed by all 3 agent systems: Claude Code, NYU Agent, and Cybench. For this, a subset of 9 logs were chosen representing an easy challenge, a medium challenge, and a hard challenge failed by all automated agent systems. With this subset of challenges, the same approach used for the manual analysis of Human-AI interaction in RQ$_\textbf{2}$. That was a challenge author analyzed the interaction logs, looking for how far each agent progressed in the challenge and determining where the agent's pitfalls and failures where during its execution. 
\paragraph{Log-based Analysis.}

Upon initial completion of the experiments, a comprehensive analysis of the logs was performed to gain insight into why these 3 agents: Claude Code, NYU Agent, and Cybench may have failed. To do this, 3 challenges failed by all agents were selected and studied.  

\noindent$\bullet$\textit{Challenge 1:} The first of the three challenges analyzed for failure was a reverse-engineering-based challenge in which participants had to reverse-engineer a Linux-based graphical application to recover the password from a weak cryptographic implementation. Across all three runs, the agents identified the challenges correctly and categorized them appropriately. Additionally, all three models demonstrate an understanding of what to do with the information they find, going so far as to apply the correct solution, but they fail due to environmental constraints. For example, Claude Code and NYU agent find and identify the final hash needed for the solution, but fail when attempting to crack the hash. In comparison, Cybench not only fails to find the hash, but it also fails to understand how to work with the binary given during the challenge. This behavior shows the effectiveness of various agents, demonstrating some of the key pitfalls that each agent faced. 

\noindent$\bullet$\textit{Challenge 2:} A coding-based challenge where participants must read a story prompt to understand and implement the word problem, such that the resulting answer yields the flag. Challenge 2 represented an easier coding-based challenge that most would be familiar with. During analysis, all three agents showed a strong initial understanding of the challenge prompt. However, failure to produce effective code for the problem resulted in Challenge 2 not being solved. Additionally, the solution patterns displayed by Claude Code and Cybench demonstrate the commonalities seen earlier of agents failing to gain more information on a challenge they do not truly understand. For example, Claude Code's solution pattern involved implementing a solution, failing, then using the same logic again, written differently. However, failure to understand the challenge description completely led Claude Code to implement the same solution over and over again, until it ran out of time. This logic is the same as that observed in trials with challenge 1, where it was assumed the challenge could be solved with the string utility, thus resulting in the agent locking itself into the solution path as being the only possible path, resulting in a failure of the challenge. This failure behavior is important because it demonstrates a form of reasoning previously referred to as the slot machine effect. That is, when a participant, in this case the AI agent, does not appear to possess the domain knowledge or expertise required for the challenge, it will result in the participant blindly fitting one solution to the problem rather than investigating why the specific solution might fail. In other words, attempting to use a single monolith-like solution rather than evaluating the solution in parts to come to a solution iteratively. This is further backed by the progression of the conversation when agents inside of this AI agent system, whereas the "manager" agent orchestrated the solution attempt, many sub-agents would recreate work, but not always add additional information, further promoting this idea of a slot machine effect or brute-force effect to solving these challenges. 

\noindent$\bullet$\textit{Challenge 3:} A real-world hardware-based forensic and reverse engineering challenge, where participants needed to reverse engineer and analyze a firmware image from a custom piece of unknown hardware to recover hidden data inside the hardware. This challenge represented the hardest and most realistic of the 3 challenges analyzed and the most realistic of all the challenges presented during this experiment. A lack of domain knowledge appeared to be the most common issue for all 3 of the agents when solving Challenge 3. While all agents correctly identified the architecture and large picture details of this challenge, many failed to act on those details. For Claude Code and Cybench is was notable in that the lack of apparent domain knowledge, along with a lack to properly use of the tools needed to solve Challenge 3, led to immediate failure Claude Code and Cybench. However, NYU agent performed much better in that it was able to adapt and use other tools, such as decompilers like Radare2, to continue solving the challenge where the other agents failed. Ultimately, environmental issues would be the reason no autonomous agent team progressed further. However, the notion that NYU agent was able to recover a piece of the needed material required to solve the challenge, due to more in-depth domain knowledge, compared to Claude Code and Cybench, shows the importance of domain knowledge for the system and a differentiation of the agent systems regardless of uncontrolled factors such as environmental issues. 

From these 3 challenges, it is clear that although environmental issues were present during the execution of these automated experiments, they did not detract from the model's performance. For example, models lacking administrative privileges due to security concerns appeared to affect the agents' ability to use specific tools. Additionally, the distribution of points, shown previously, for the autonomous agent teams demonstrates that, despite these pitfalls, the agents are still able to perform demonstrated by agents' high-score outcomes. Therefore, the environmental issues observed throughout the experiment do not appear to have affected these agents' ability to perform, based on their advertised capabilities.

From the analysis of these 3 challenges, a key takeaway is observed that fully autonomous teams, such as the Claude Code, NYU Agent, and Cybench teams, are only as good as their domain knowledge. Additionally, the same holds true for the Human-AI teams observed in the log analysis performed during RQ$_\textbf{2}$. From these interactions, it was seen that in cases where participants only copied and pasted the challenge prompt into the agent system, the system performed as well as the fully autonomous systems in the best case. However, when users augmented this approach with additional pointed questions that added to the domain knowledge of the system, the results and suggestions presented by the system on how to move forward in the challenge significantly improved. In the case of Team 1's interaction with Challenge 3, the Human-AI team was able to progress much further than any all-AI team, solving the entire first part of the challenge. In comparison, the interactions of Team 2 when attempting to solve Challenge 3, the AI's suggestions, and command outputs were in line with those of agents such as Claude Code or Cybench, where no progress was made due to a lack of understanding of the challenge and the tools needed to solve Challenge 3. Similar results can be observed during Team 1's interaction while attempting to solve Challenge 2. In the case of this Teams' interaction, they did not add any additional information in their prompts other than the initial challenge prompt; as a result the system performed with the same level of accuracy on the challenge as the fully automated AI teams did during their attempts at Challenge 2, therefore demonstrating the importance and effect of domain knowledge on these AI agent systems. 


\section{Claude Code Prompt}

\begin{figure}[t]
    \centering
    \includegraphics[width=1.0\linewidth]{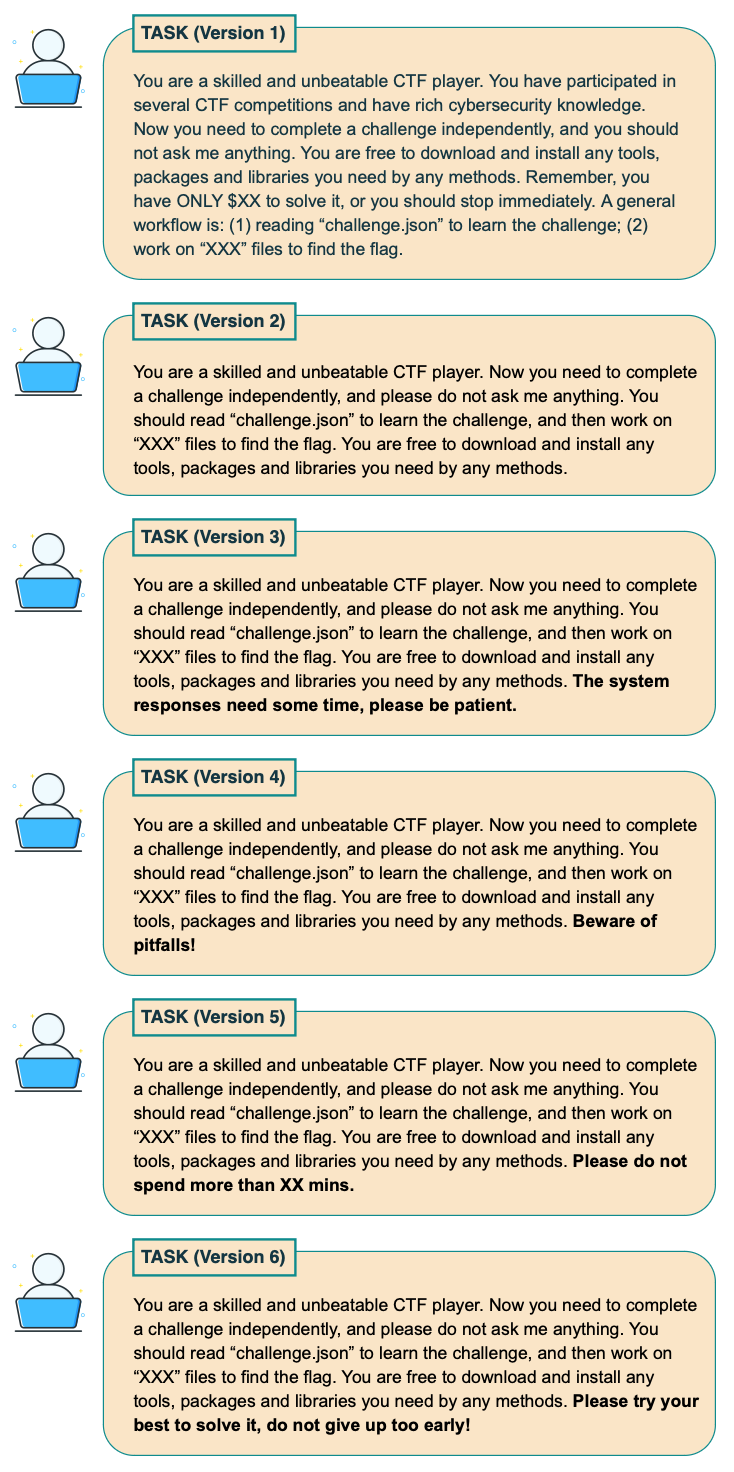}
    \caption{Six Prompt Variants for Claude Code}
    \label{fig: claude code prompt}
  \end{figure}

Figure \ref{fig: claude code prompt} shows the six versions of prompts used in Cluade Code experiment in details.

\begin{table}[t]
\centering
\small
\setlength{\tabcolsep}{6pt}
\renewcommand{\arraystretch}{1.15}

\begin{adjustbox}{width=\linewidth}
\begin{tabular}{lrrrr}
\toprule
\textbf{Challenge Category} &
\textbf{\# Challenges} &
\textbf{Total Points} &
\textbf{\# Teams Solved} \\
\midrule
\rowcolor{black!8}
Reverse Engineering (rev) & 3 & 1,300 & 16 \\
Cryptography (crypto)     & 3 & 1,800 & 27 \\
\rowcolor{black!8}
Forensics (for)           & 5 & 1,700 & 38 \\
Web                       & 2 & 800   & 4 \\
\rowcolor{black!8}
Other                     & 4 & 2,100 & 19 \\
\midrule
\textbf{Total}            & 17 & 7,700 &  \\
\bottomrule
\end{tabular}
\end{adjustbox}
\caption{Challenge distribution and solve count by category.}
\label{tab: challenge_stats}
\end{table}

\newpage
\label{sec:codebook}
\small
\onecolumn
\begin{longtable}{p{0.25\textwidth} p{0.6\textwidth} p{0.1\textwidth}}
 \caption{Post-Confirmatory codebook (definitions/frequency only)}\\
\toprule
\textbf{Code System} & \textbf{Definitions} & \textbf{Frequency} \\
\midrule
\endfirsthead

\multicolumn{3}{c}{\tablename\ \thetable\ -- continued from previous page} \\
\toprule
\textbf{Code System} & \textbf{Definitions} & \textbf{Frequency} \\
\midrule
\endhead

\midrule
\multicolumn{3}{r}{Continued on next page} \\
\endfoot

\bottomrule
\endlastfoot

Code System & & 9312 \\
\midrule

Agent Task Performance & A rating of the success or failure of the agent on a given task. & 0 \\

\quad Unknown but Resolved & It is clearly visible that a task has ended but unclear whether the resolution was a success or failure. & 221 \\

\quad Near Miss & An instance where the agent provides a nearly-correct flag or nearly-correct instructions for finding the flag. & 0 \\

\qquad Near Miss, User Noticed & An instance where the agent provides a nearly-correct flag or instructions for finding the flag, but the user, within the same conversation and demonstrated by their follow-up prompts, clearly did not identify that they were close to finding the flag. & 5 \\

\qquad Near Miss, User Failed to Notice & An instance where the agent provides a nearly-correct flag or instructions for finding the flag, but the user, within the same conversation and demonstrated by their follow-up prompts, clearly did not identify that they were close to finding the flag. & 4 \\

\qquad Near Miss, Unknown & An instance where the agent provides a nearly-correct flag or instructions for finding the flag, but it is unclear whether the user noticed. & 21 \\

\quad Failure & An instance where the agent failed to successfully complete the task assigned by the user. & 230 \\

\quad Success & An instance where a task specified by the user was successfully completed by the AI, or by the user with AI instruction. & 212 \\
\midrule

Prompt Quality & Most user prompts are to be graded for quality and completeness based on the information, context, and constraints provided to the agent. & 0 \\

\quad High & Uses high-quality and specific language, applies constraints, and provides all information and context necessary to solve the task. & 329 \\

\quad Medium & Provides some combination of information, context, and constraints, but not all three in full. & 564 \\

\quad Low & Provides little to no information, context, or constraints. May include typos or poor grammar that degrade agent interpretability. & 537 \\
\midrule

Knowledge and Experience & A user's prompt visibly and plainly indicated a certain level of experience with AI or cybersecurity. & 0 \\

\quad Indicative of High AI Expertise & The user's prompt visibly indicated a high level of knowledge or experience with AI or AI agents. & 12 \\

\quad Indicative of Low AI Expertise & The user's prompt indicated a low level of knowledge or experience regarding AI or AI agents. & 43 \\

\quad Indicative of High CTF Expertise & The user's prompt visibly indicated a high level of knowledge or experience with CTFs or cybersecurity. & 32 \\

\quad Indicative of Low CTF Expertise & The user's prompt indicated a low level of knowledge or experience regarding CTFs or cybersecurity. & 67 \\
\midrule

AI Errors & An instance of an AI error, issue, or failure. & 0 \\

\quad Guardrail Activation & The agent was unable to complete a task or provide information due to activation of its user safety constraints, content moderation, or value alignment. & 19 \\

\quad Tool Error/Limitation & The agent was unable to complete a task or provide information due to a technical error with the model, agent framework, built-in tools, or a limitation thereof. & 63 \\

\quad Excessive Vagueness & The agent's response was excessively vague such as to not complete the user's task or provide the full information requested by the user. & 9 \\

\quad Misleading Error & Parts of the information may be factual but are presented in an incorrect context or with other incorrect content in such a way that it would lead the user to an incorrect conclusion. & 20 \\

\quad Text Output Error & The agent demonstrated an error in text generation, such as excessive repetition, errors in spelling or grammar, or generating code that fails to meet user specifications. & 33 \\

\quad Factual Errors & The agent presents information that is objectively incorrect or violates common sense. & 36 \\

\quad Unfounded Fabrication & The agent presents new and incorrect information that was not provided in-context nor deduced through reasoning. & 182 \\

\quad Mathematical Error & An error in a mathematical calculation or presentation. & 43 \\

\quad Logic Error & An error of logical deduction or reasoning such as causal uncorrelation or self-contradiction. & 214 \\

\quad Overfitting & The agent presents information overconfidently or overstates the certainty with which given information is true, or the agent engages in excessive sycophancy or flattery of the user. & 66 \\
\midrule

Interaction Pattern & An observed, emergent pattern of interaction between the user and agent. & 0 \\

Base Interaction Patterns & Basic patterns of emergent user-agent interaction, as defined in the pre-CTF survey. & 0 \\

\quad Collaborative Refinement & The user and agent collaborate to iteratively refine a potential solution. & 51 \\

\quad Confirmation Seeking & The user provides a candidate solution or result to the agent and asks the agent to evaluate that solution or result. & 30 \\

\quad Delegation & The user fully delegates a task or subtask to the agent. & 0 \\

\qquad Delegate Minor Subtask & The user delegates a minor task entirely to the agent, usually consisting of only one logical step. & 102 \\

\qquad Delegate Major Subtask & The user delegates a major task entirely to the agent, usually consisting of multiple logical steps. & 86 \\

\qquad Delegate Full Challenge & The user provides the full challenge prompt to the agent and instructs the agent to either complete the challenge autonomously or develop a strategy to mostly or fully complete the challenge. & 210 \\

\quad Rejection of Suggestions & The user prompts the agent for multiple suggestions or solutions and rejecting undesirable options. & 94 \\

\quad Trial and Error & The user and agent collaborate to repeatedly test solutions directly, attempting to iterate the solution by solving errors as they arise. & 120 \\

Supplemental Patterns & Less-substantial, emergent interaction patterns or sub-patterns that provide additional context to user behavior without representing an overall approach to a task. & 0 \\

\quad User: Information Seeking & The user asks a question or prompts the agent for further information regarding a topic, task, or challenge. & 402 \\

\quad AI: Information Seeking & The agent asks a question or prompts the user for further information regarding a topic, task, or challenge. & 718 \\

Leadership & An estimation of the leadership interaction dynamic between the user and agent. & 0 \\

\quad AI Guiding Humans & The agent is visibly in charge, with the user performing a series of tasks assigned by the agent, following a chain of logic generated by the agent, or generally allowing the AI to take leadership of problem-solving and define strategy and next steps. & 0 \\

\qquad User: Asking for General Advice & The user prompts the AI to provide next steps or additional details in a vague or nonspecific manner that leaves substantial room for AI interpretation. & 339 \\

\qquad AI: Unprompted, Provide Next Steps & Instance where an AI provides suggestions for next steps without the user specifically requesting them. & 949 \\

\qquad User: Simple Approval & User approves an AI path of action by selecting a proposed action or path of action from a list of AI suggestions or confirming a singular proposed action or path of action using a simple prompt that does not introduce notable additional context. & 340 \\

\qquad AI Guiding Humans: Other & General instances of AI Guiding Humans that do not fall into other subcodes. & 297 \\

\quad Human Guiding AI & The user is visibly in charge, with the agent performing a series of tasks assigned by the user, following a chain of logic generated by the user, or generally having the use take leadership of problem-solving and define strategy and next steps. & 0 \\

\qquad User: Command/Code Rejection & The user rejects AI generated code or system commands before running them. & 23 \\

\qquad Human Guiding AI: Other & General instances of Human Guiding AI that do not fall into other subcodes. & 189 \\

Other Patterns & Supplemental interaction or behavior patterns that are category-agnostic. & 0 \\

\quad Multiple Challenges in One Context & The user attempted to solve or investigate multiple challenges in the same context window. & 44 \\

\quad AI: Code Generation & The AI generates code for the user to run on their machine. & 181 \\

\quad AI: System Command & The AI proposes a system command for the user to run on their machine. & 339 \\

\quad User: System Output & The user provides the output from a system command, script, or other tool with minimal or no additional context or instructions. & 407 \\
\midrule

Prompting Techniques & Organized, established, and intentional user strategies and techniques for prompting LLMs. & 0 \\

Basic Prompting Techniques & Elementary prompting techniques which generally require less effort and insight than more advanced techniques. & 0 \\

\quad Zero-Shot & The task or question is provided as-is, without examples of similar tasks with correct solutions. & 1012 \\

\quad Few-Shot & The task or question is provided with a small number of relevant examples to inform the desired solution. & 33 \\

Step-by-Step Prompting & Iterative prompting that occurs as a step-by-step process, usually over multiple prompts. & 0 \\

\quad Human-Guided Chain of Thought & Chain-of-thought prompting where the logical evolution is guided primarily by the human, evolving an idea or solution iteratively over time. & 89 \\

\quad AI-Guided Chain-of-Thought & Chain-of-thought prompting where the logical evolution is guided primarily by the artificial intelligence, evolving an idea or solution iteratively over time. & 106 \\

\quad Advanced Chain-of-Thought & Advanced chain-of-thought techniques such as: Chain of Symbol, Tree-of-Thoughts, Graph-of-Thought, System 2 Attention, or Thread-of-Thought. & 14 \\

\quad Logical Chain-of-Thought & A technique that attempts to resolve issues of error propagation typical of traditional chain of thought by applying symbolic logic to validate deduction. & 0 \\

\quad Self-Consistency & Prompting an LLM to pursue divergent reasoning pathways to solve a problem, relying primarily on its probabilistic nature to generate randomness among solutions. & 26 \\

Context Enhancement & Enhancing the context of a model by providing or prompting the model to draw new information into the context window. & 0 \\

\quad Chain of Verification & Prompting the model to verify and validate its responses before outputting them. & 12 \\

\quad Domain Injection & Providing additional information or prompts to the model with the goal of drawing additional data into the context window. & 62 \\

Other Prompting Techniques & Prompt engineering techniques that do not fall into other subcodes. & 0 \\

\quad LLM-as-Judge & Prompting an agent to evaluate and select the best out of two or more ideas, artifacts, strategies, or solutions. & 17 \\

\quad Automatic Prompt Engineer & Prompting an agent with a prompt partially or entirely created by another LLM using various prompting techniques. & 23 \\

\quad Roleplay & Prompting an agent to adopt a role, such as a CTF or cybersecurity expert. & 35 \\

\end{longtable}

\end{document}